\documentclass[a4paper,fleqn]{cas-sc}

\usepackage[authoryear]{natbib}
\usepackage{soul}
\usepackage{subcaption}
\usepackage{graphicx}
\graphicspath{{./}{figs/}}

\def\tsc#1{\csdef{#1}{\textsc{\lowercase{#1}}\xspace}}
\tsc{WGM}
\tsc{QE}
\tsc{EP}
\tsc{PMS}
\tsc{BEC}
\tsc{DE}

\begin{document}
\let\WriteBookmarks\relax
\def\floatpagepagefraction{1}
\def\textpagefraction{.001}

\ExplSyntaxOn
\cs_gset:Npn \__first_footerline:
  { \group_begin: \small \sffamily \__short_authors: \group_end: }
\ExplSyntaxOff 
\shorttitle{Experimental and numerical investigation of the flow through packed beds}

\shortauthors{Patil, Gorges, L\'opez~Bonilla, Stelter, and van Wachem}

\title [mode = title]{Experimental and numerical investigation to elucidate the fluid flow through packed beds with structured particle packings}

\author[affil1]{Shirin Patil}
\author[affil2]{Christian Gorges}
\author[affil1]{Joel L\'opez~Bonilla}
\author[affil1]{Moritz Stelter}
\author[affil1]{Frank Beyrau}
\author[affil2, cor1]{Berend van~Wachem}[type=editor,
      orcid=0000-0002-5399-40750,
      ]
\cortext[cor1]{Corresponding author: Berend van Wachem, Email: berend.vanwachem@ovgu.de}

\address[affil1]{Laboratory of Technical Thermodynamics,
Otto-von-Guericke-Universit\"{a}t Magdeburg,\\ Universit\"atsplatz 2, 39106 Magdeburg, Germany}
\address[affil2]{Chair of Mechanical Process Engineering,
Otto-von-Guericke-Universit\"{a}t Magdeburg,\\ Universit\"atsplatz 2, 39106 Magdeburg, Germany}

\begin{abstract}
The present paper presents an experimental and numerical investigation of the dispersion of the gaseous jet flow and co-flow for the simple unit cell (SUC) and body centered cubic (BCC) configuration of particles in packed beds. The experimental setup is built in such a way, that suitable and simplified boundary conditions are imposed for the corresponding numerical framework, so the simulations can be done under very similar conditions as the experiments. Accordingly, a porous plate is employed for the co-flow to achieve the uniform velocity and the fully developed flow is ensured for the jet flow. The SUC and BCC particle beds consist of 3D-printed spheres, and the non-isotropy near the walls is mostly eliminated by placing half-spheres at the channel walls. The flow velocities are analysed directly at the exit of the particle bed, for both beds over 36 pores for the SUC configuration and 60 pores for the BCC configuration, for particle Reynolds numbers of 200, 300, and 400. Stereo particle image velocimetry (SPIV) is experimentally arranged in such a way, that the velocities over the entire region at the exit of the packed bed are obtained instantaneously. The numerical method consists of a state of the art immersed boundary method with adaptive mesh refinement. The paper presents the pore jet structure and velocity field exiting from each pore for the SUC and BCC packed particle beds. The numerical and experimental studies show a good agreement for the SUC configuration for all flow velocities. For the BCC configuration, some differences can be observed in the pore jet flow structure between the simulations and the experiments, but the general flow velocity distribution shows a good overall agreement. The axial velocity is generally higher for the pores located near the centre of the packed bed than for the pores near the wall. In addition, the axial velocities are observed to increase near the peripheral pores of the packed bed. This behaviour is predominant for the BCC configuration as compared to the SUC configuration. The velocities near the peripheral pores can become even higher than at the central pores for the BCC configuration. It is shown that both the experiments as well as the simulations can be used to study the complex fluid structures inside a packed bed reactor.

 \end{abstract}




\maketitle

\section{Introduction}

Packed bed reactors, especially with gaseous flows, have wide-ranging engineering applications. For example, in the food industry (e.g., bioreactors for dairy products production or coffee roasters), basic materials' industry (e.g., shaft kilns to produce lime or dolomite), energy sector (e.g., production of synthesis gas from biomass), to name just a few examples. In such applications, there are multi-phase interactions, which are governed by physical phenomena such as mass transfer, heat transfer and the fluid flow through the packed bed. The construction of fixed packed bed reactors usually relies on simplifying assumptions, such as plug flow \citep{Eppinger2011, Dixon2020} or empirical correlations, such as the Ergun equation \citep{Ergun1952}. However, such assumptions can give rise to erroneous predictions, principally for small tube to particle diameter ratios, as a result of wall effects \citep{Nijemeisland2001}. The distribution of flow within the reactor and the development of the velocity field in the freeboard above the interface can significantly affect the overall process. Thus, the local flow behaviour within the bed is a crucial parameter for optimising fixed packed bed reactor systems.

Currently, there is limited experimental information available on the local fluid flow structure within packed particle beds, as accessing the particle interstices experimentally is highly challenging without affecting the flow. Nevertheless, a variety of experimental techniques have been carried out to study different aspects of the fluid flow through packed beds in different configurations. Probe-based techniques, such as hot wire anemometry \citep{Alshammari2023HWA} or electrochemical micro-probes \citep{Seguin1998,Bu2015}, are intrusive but fast frequency techniques ($\sim$ 100 Hz) that have been used to measure the instantaneous local flow velocity within the interstitial spaces of a packed bed. The first is limited to applications with gas flows, while the second with liquid flows. Laser Doppler velocimetry \citep{giese1998measured,CHEN2007LDV} is a semi-intrusive optical technique but, as the previous technique, allows measuring the instantaneous velocity in a single point, which limits the achieved understanding of the fluid flow phenomena. Magnetic resonance imaging (MRI) allows determining the three components of the interstitial velocity of liquid flows in the interparticle spaces of different packed beds, as random packed beds with Ballotini particles \citep{sederman1997magnetic} and simple unit cell (SUC) packed bed \citep{suekane2003inertial}. This technique is highly expensive and few works have been performed with gas flows \citep{Nguyen2005MRI}.

Particle image velocimetry (PIV) is a technique that can measure two or three components of the instantaneous velocity of a fluid seeded with particles. The particles are illuminated by laser light, and the images are recorded with synchronized cameras \citep{Raffel2018}. This technique has different configurations that have been widely used to characterize the flows associated with packed bed. Several works are performed using planar PIV on transversal planes to packed beds of solid spheres using the refractive index matching (RIM) method \citep{wiederseiner2011refractive, Haam2000RIM}, which helps to access to the interstitial spaces minimizing the distortions in the captured images, by having the similar refractive index for the fluid and spheres, but it is limited to liquid flows. The following works used this method to evaluate liquid fluids at low particle Reynolds number, $\mathrm{Re}_\mathrm{p}$, in random packed beds with monodisperse spheres. \cite{huang2008optical} reported for $\mathrm{Re}_\mathrm{p}$ = 28 that the velocity in the pore increases with porosity size and the velocity asymmetries around the spheres are influenced by the fluid inertia. \cite{patil2013flow} have studied $\mathrm{Re}_\mathrm{p}$= 4 in a low aspect ratio packed bed and evaluated three different locations (vertical planes), finding that the flow structures become less ordered and the dynamic range of velocities increases from near wall towards the midplane. Furthermore, the following works have used the RIM method to evaluate the turbulence intensity in random packed beds with monodisperse spheres. \citep{patil2013turbulent} have applied time resolved planar PIV to study turbulent flow with $\mathrm{Re}_\mathrm{p}$ ranging from 418 to 3964, where they identify repetitive patterns in the pore spaces and demonstrate that most of the turbulent measures become independent of $\mathrm{Re}_\mathrm{p}$ beyond $\mathrm{Re}_\mathrm{p}=2800$. \cite{khayamyan2017measurementsPIV} have studied turbulent and transitional flow in a randomly monodisperse packed bed at $\mathrm{Re}_\mathrm{p}$ ranging from 20 to 3220. They found that when the $\mathrm{Re}_\mathrm{p}$ increases, the magnitude of velocities increases, the dynamic range of velocities decreases, but the flow is more disordered, predominately in low velocity areas. Also, they define flow regimes, such as the regime of stokes to inertial transition for $\mathrm{Re}_\mathrm{p}$ from 40 to 250, inertia to turbulent transition for $\mathrm{Re}_\mathrm{p}$ from 250 to 1500 and turbulent flow from $\mathrm{Re}_\mathrm{p}$=1500, where the velocity fluctuation becomes independent of $\mathrm{Re}_\mathrm{p}$. The same group, applied Stereo PIV with RIM to study the longitudinal and transversal dispersion of transitional and turbulent flows throughout the packed bed \citep{khayamyan2017transitionalSPIV}. Further works also applied tomographic PIV \citep{larsson2018tomographic} or time resolved planar PIV \citep{nguyen2018time, nguyen2019experimental, nguyen2021experimental} to study liquid flows in packed beds. Planar PIV without RIM, but with optical access, has been used in some works to measure the velocity fields of liquid flows \citep{blois2012quantifying} and gas flows. \cite{velten2023flow} and \cite{Neeraj2023}, both from the same group, measure the velocity fields of gas flow, at $\mathrm{Re}_\mathrm{p}$ from 200 to 500, in some pores and the freeboard above a packed bed of spheres arranged in body centered cubic (BCC) packing. They evaluates the influence of the number of layers on the flow above the bed, where it was found that from 11 layers, the surface flow becomes independent of the number of layers, but 21 layers minimize the influences from the surroundings. The velocity profiles above the bed have been analysed, observing that a non-periodic porosity distribution near the wall creates a channelling effect, which  leads to high velocity jets near the wall. Also, the velocity profiles closer to the top of the bed show clearly the jets from the inter-particle spaces. About the influence of the $\mathrm{Re}_\mathrm{p}$, the averaged flow structures are not influenced, but at higher $\mathrm{Re}_\mathrm{p}$, the presence of recirculations around the sphere is more evident, the flow structures fluctuate more, which could derive in asymmetric averaged velocity fields. These works are still limited to the study of pores that have optical access, but to access to the pores behind the spheres in the gas phase, where RIM is not applicable, planar PIV with image correction methodology based on ray tracing PIV (RT-PIV) has been proposed to correct the optical aberrations from the transparent spheres \citep{martins2018ray}. This technique has been validated to be used in a BCC packed bed by \cite{velten2024ray}, who show that RT-PIV still has a limited field of view, is very sensitive to geometric parameters, and has difficulties in the illumination.

Next to observing the physical behaviour of a system, experimental studies can also assist in the validation of numerical models \citep{wood2015comparison}. Simulations of fixed packed particle beds employing computational fluid dynamics (CFD) assume a vital role in predicting and regulating flow and process parameters, both nowadays and in the near future. Several simulation methods can be used to simulate the fluid dynamics inside fixed bed reactors. The Lattice-Boltzmann method (LBM) is often used due to its good scalability and efficiency, but it has drawbacks for dense particle packings and high Reynolds number turbulent flows with additional heat or mass transfer \citep{Mantle2001, Manz1999, Sullivan2005, Dixon2020}. Another common method is to solve the Navier-Stokes equations as a single-phase flow on body-fitted finite volume meshes. This method is accurate, depending on the complexity and structure of the used numerical mesh, but it can be computationally very expensive, especially for complex particle shapes \citep{Eppinger2011, Robbins2012, Yang2013}.

A third option is the immersed boundary method (IBM), which does not require the fluid mesh to conform to the surfaces of the particles. This makes IBM significantly less computationally expensive than body-fitted meshes, while still maintaining a good accuracy. The IBM is a numerical method for simulating fluid flow around complex geometries, such as the particles in a fixed bed reactor. The IBM was first introduced by \cite{Peskin1972} for the simulation of fluid-structure interactions in heart valves, and has since been extended to a wide range of applications. The IBM uses an Eulerian framework for the discretization of the fluid domain and a Lagrangian marker framework for the representation of the particle surface. This means that the fluid mesh does not conform to the surfaces of the particles, which simplifies mesh generation and reduces computational costs. In the continuous forcing IBM, also known as the smooth IBM, the particle surfaces are represented by source terms in the Navier-Stokes equations. These source terms are spread across several fluid cells at each side of the particle surface \citep{Peskin1972}. The no-slip boundary condition at the particle surface is imposed by requiring that the fluid velocity at the Lagrangian markers match the desired velocity at the surface.

In the realm of fixed bed reactor simulation, the IBM has gained notable prominence in recent years. A study conducted by \cite{Gorges2024} delved into the comparison between two IBM approaches: the smooth and blocked-off IBM techniques. These methods were applied to simulate a fixed packed bed reactor consisting of spherical particles arranged in a BCC particle packing structure. The investigation involves an examination of velocity fields in vertical planes above the fixed packed bed, with a focus on varying Reynolds numbers. To bolster their findings, the researchers relied on experimental inline PIV measurements as a foundational benchmark for evaluating the efficacy and accuracy of the two IBM approaches.

Another notable contribution to this field is from \cite{Yuan2019}, who applied the IBM in conjunction with local adaptive meshing techniques. This combination has been employed to simulate fixed bed reactors containing particles of various shapes. Their study entails a thorough comparison of the predicted pressure drop across the bed and local heat transfer with empirical correlations corresponding to these parameters.

Furthermore, \cite{Lovreglio2018} has extend the exploration of fixed bed reactor dynamics. The study involves a comparative analysis between results obtained through MRI and those generated by employing a proprietary CFD code built with an IBM framework. The primary focus was to elucidate the structural aspects and hydrodynamics of fixed beds comprising spherical particles.

In the current paper, the focus is to analyse the fluid velocity field of the outflow close to the exit of a packed bed, and to study the dispersion of a fluid jet flow through the packed bed for different flow conditions, from $\mathrm{Re}_\mathrm{p}$=200 to $\mathrm{Re}_\mathrm{p}$=400. In the present study, the packed bed is arranged in two structured configurations: simple unit cell (SUC) and body centered cubic (BCC). The velocity fields, predominantly the axial vector, are measured by the stereo-PIV (SPIV) technique and provide experimental data to compare with numerical results obtained by the smooth IBM~\citep{Cheron2023a}.

Previous studies have performed planar PIV measurements at maximum three locations (vertical planes) in the outflow and under the presence of one or two interstitial spaces at the exit of packed bed \citep{suekane2003inertial,velten2023flow,Neeraj2023,velten2024ray,Gorges2024}. This approach cannot continuously resolve the velocity field over the entire plane, as it misses the information from the spatial gaps. The novelty of this work is the study of jet dispersion throughout a wide packed bed, with multiple interstitial spaces, by measuring simultaneously the three components of velocities over the entire plane at the exit of the packed bed, which is possible with the SPIV technique. The experimental data used to validate numerical simulation reproduces, as much as possible, the simplified and well-defined boundary conditions, which are required in a numerical calculation. The experimental particle packed beds have been 3D-printed with a high dimensional precision technique, allowing an accurate positioning of all the spheres in the experimental and numerical setup, the jet flow originates from a fully developed flow in the bottom centre from  a pipe, and the co-flow has a uniform velocity, ensured by using a sufficiently thick porous plate. The printed packed particle bed also ensures the uniform porosity within the bed, specially at the walls interface, where half spheres instead of full spheres have been printed \citep{Bu2015}. This is critical to minimize the wall flow and its influence on the surface velocity field \citep{Gorges2024,Neeraj2023}.

\section{Experimental setup and methodology} \label{sec:Experimental setup and methodology}

In this section, we discuss the experimental setup for the SUC and BCC packed particle bed arrangements. The optical setup and the methodology of image processing for SPIV is also outlined and discussed. 

\subsection{Experimental Setup} \label{subsec:Experimental setup}

\begin{figure}[!t]
	\centering
        {\includegraphics[width=16cm]{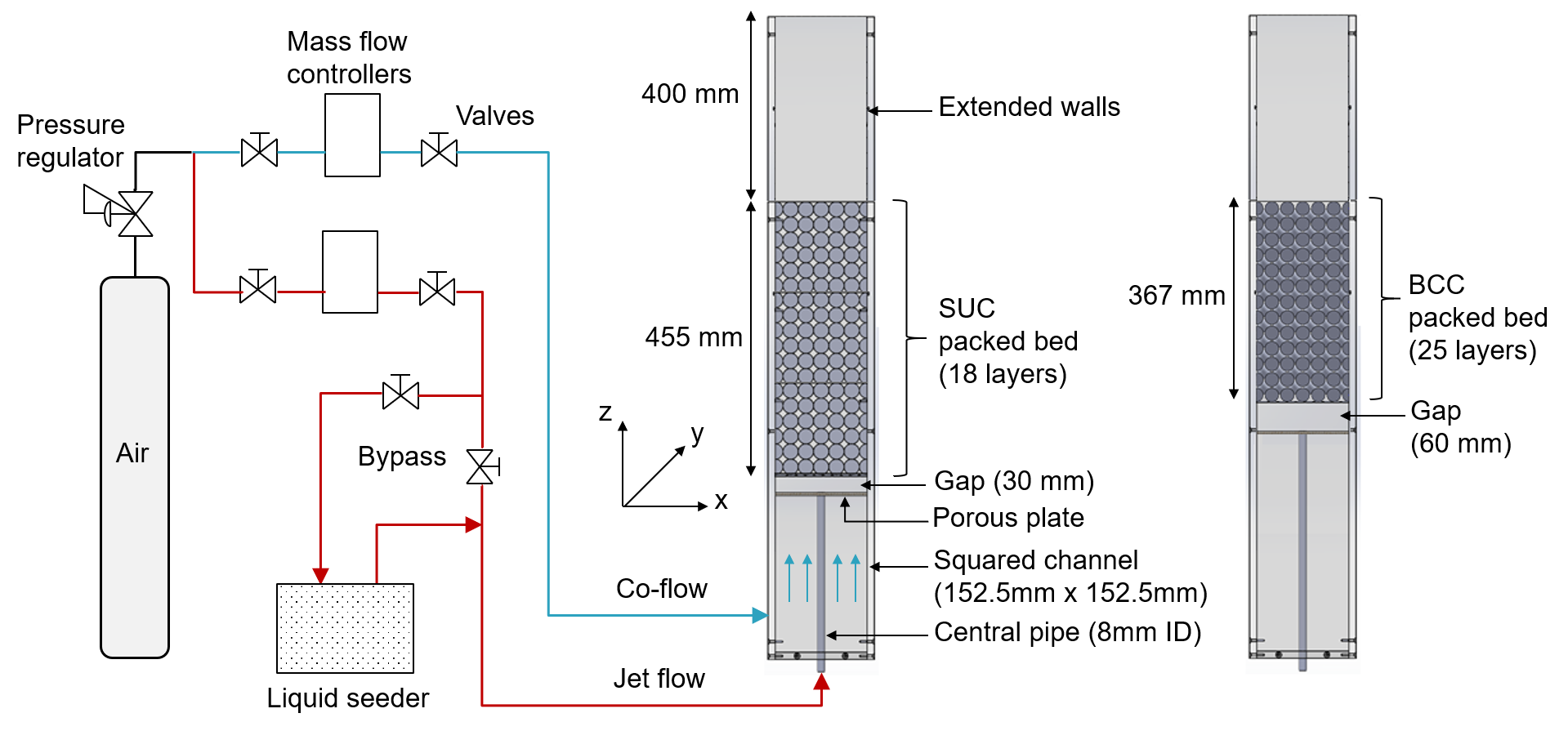}}
	\caption{Experimental packed bed setup with the SUC and BCC arrangements and schematic gas routing.}
	\label{fig:expt-setup}
\end{figure}

Figure \ref{fig:expt-setup} shows the experimental packed bed setup, including the SUC and BCC particle arrangements. The setup consists of a jet flow in the centre of the bed with a concentric co-flow that interacts with the packed bed from bottom to top. Both flows consist of synthetic air. The main components of the setup are the square channel, the central pipe for the central jet flow, the porous plate at the bottom for the co-flow, and the packed bed. The squared channel conducts the co-flow and has a cross-section of 152.5 mm x 152.5 mm with walls made of transparent acrylic (PMMA). The central pipe is made from stainless steel and conducts the jet flow, seeded with particles. It has an inner diameter of 8 mm, an outer diameter of 12 mm and a length of 380 mm, so that the inner flow is fully developed before exiting the pipe. The packed beds are 3D-printed (made from Nylon PA-12) and consist of uniform spheres placed periodically in SUC or BCC arrangements. The 3D printing and design details are described in Section \ref{subsec_3dprinting}. Before the co-flow interacts with the spheres of the packed bed, it is homogenized by passing through a bronze porous plate (Siperm, B40 with 10 mm thickness) that fully covers the channel cross-section. A uniform velocity profile across the entire section is desirable, as it provides a known and homogeneous boundary condition for the simulation work. A circular hole has been drilled exactly in the middle of the porous plate to hold the central pipe for the jet flow, keeping it centred in the square channel. As shown in Figure \ref{fig:expt-setup}, the exit of the pipe is flush fitted with the end surface of the porous plate, such that the uniform co-flow and the jet flow exit at the same plane. The packed bed is held 30 mm above porous plate for SUC configuration or 60 mm, for BCC configuration, using a 10 mm step on the inside of the channel walls. This creates a gap between the exit of jet flow and the inlet of packed bed, which allows some room for jet flow to disperse before interacting with the layers of packed bed. After emerging from the bronze plate plane, both the co-flow and jet flow enter the packed bed and continue to flow upwards, crossing all the layers of the packed bed. To ensure that the flow at the exit of the packed bed is not affected by any kind of perturbations from the surrounding, the walls of the square channel are extended to 400 mm beyond the last layer of the packed bed. This provides a well-defined boundary condition for the simulation part of this study. In this work, the SUC packed bed consists of 18 layers of spheres, and the BCC of 25 layers. For these numbers of layers, the seeding particles from the central jet are dispersed over the entire region of the packed bed. Also, according to \citet{Neeraj2023}, in a BCC configuration, the pore jet velocities do not show much variation after 21 BCC layers. The total length of the SUC and BCC packed bed is 455 mm and 367 mm, respectively.

The required airflow is supplied from compressed air cylinders, and separate mass flow controllers are used to control the co-flow  (Bronkhorst, Mass-Stream 6371, maximum flow 380 $\mathrm{lpm}$ air) and the jet flow (Bronkhorst, El-Flow, maximum flow 80 $\mathrm{lpm}$ air). The seeding for the SPIV measurements is supplied along with the jet flow. A Lavision 'Aerosol generator' is used with Di-Ethyl-Hexyl-Sebacat (DEHS) to seed droplets into the jet flow for SPIV experiments. A liquid seeder, rather than a solid particle seeder, is used to keep the extended walls transparent and to not cloud the view of the cameras. As shown in Figure \ref{fig:expt-setup}, a bypass is used to regulate the fraction of the jet flow passing through the aerosol generator to provide control over the seeding density. In the current configuration of the setup, the co-flow cannot be seeded, as the porous plate does not allow the droplets to pass further downstream.

\subsection{The 3D-printed packed bed}\label{subsec_3dprinting}

\begin{figure}[!t]
	\centering
	\subcaptionbox{SUC\label{fig:SUCa-structure}}{\includegraphics[width=12cm]{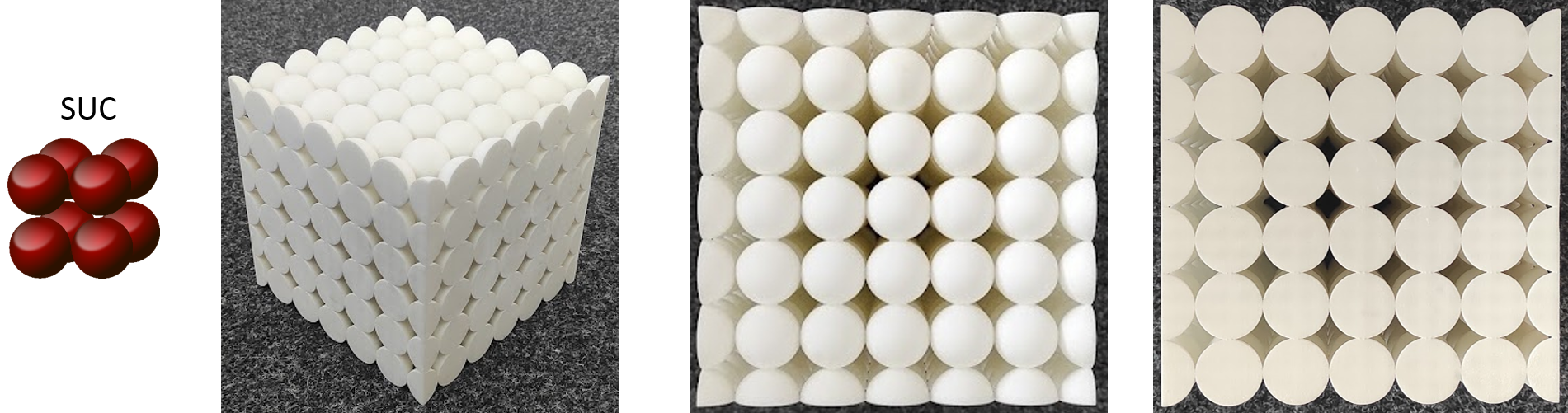}}\\
	\subcaptionbox{BCC\label{fig:BCCb-structure}}{\includegraphics[width=12cm,clip = 0cm 0cm 0cm 0.0cm]{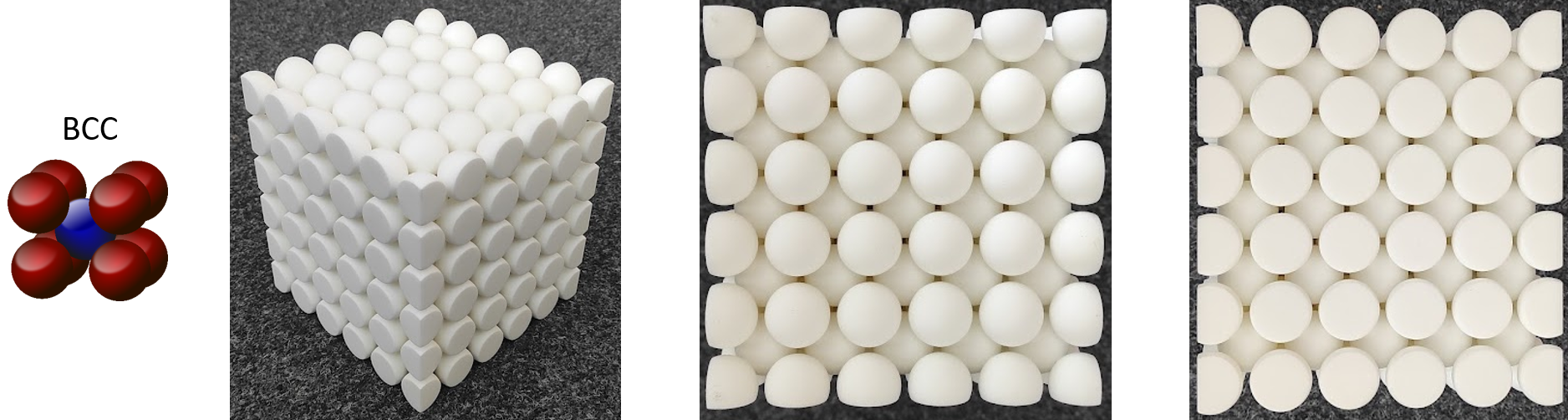}}
	\caption{3D-printed packing unit for the: (a) SUC configuration and (b) BCC configuration. Isometric, top, and side view (left to right).}
	\label{fig:SUC-BCC-structure}
\end{figure}

This section outlines the fabrication of the packed particle bed. As mentioned previously, two packed bed configurations, SUC and BCC, are investigated. These packed beds are 3D-printed using the selective laser sintering (SLS) method and the material is Nylon PA-12. To ease the handling of the packing, printing is carried out in separate packing units, which are shown in Figure \ref{fig:SUC-BCC-structure}. The spheres for both configurations have a diameter of 25.5 mm. For instance, one layer of SUC unit has 36 spheres and a unit consists of 6 layers of spheres in axial ($Z$) and lateral ($X$,$Y$) directions (see Figure \ref{fig:SUC-BCC-structure}a). However, along the $X$ and $Y$ directions, each layer of SUC unit consists of 5 full spheres and 2 half spheres, each towards the end of the packing. This boundary condition has been used before by \citet{Bu2015}, and the intention is to uniform the porosity of the packed bed near the wall, minimizing the wall flow and the channelling effects due to bigger pores near the wall which influence the superficial flow above the bed \citep{Neeraj2023, Gorges2024}. In total, the configuration has 18 layers of SUC, thus the packed bed consists of a total of 648 spheres. 
The BCC unit has two types of layers, a full layer, that corresponds to a layer which extends until the edges of a unit and a weak layer which is defined as the layer between two full layers. One full layer of a BCC unit consists of 36 spheres, while a weak layer has 25 spheres. A BCC unit consists of 11 layers, 6 full layers and 5 weak layers, along the axial ($Z$) direction (see Figure \ref{fig:SUC-BCC-structure}b). Similar to the SUC particle packing, to keep uniform the porosity near the wall, along the lateral $X$ and $Y$ direction, each full layer of BCC unit  has 5 full spheres and two half-spheres near the periphery, while each weak layer has 5 full spheres. The BCC particle packing begins and ends with a full layer. Additionally, each unit of packed bed is 3D-printed in such a way that it can be easily moved and placed over other units. In the case of the BCC packing, it is not straightforward to move and place one unit over another, hence a bridge unit, consisting of 1 full layer and 2 weak layers, is 3D-printed and added. This makes a total of 768 spheres in the BCC packed particle bed configuration. 

The pore volume fraction (void fraction) for the fluid flow within the SUC and BCC particle packings is 0.471 and 0.296, respectively. However, the theoretical volume fraction for SUC and BCC packings are 0.48 and 0.32, respectively. It is noted that the volume fractions for both configurations are slightly lower than its corresponding theoretical value. This is because in 3D printing, the contact point between each sphere has a finite size. The actual volume fraction for BCC is reduced further, as in that configuration, each sphere is surrounded by more spheres as compared to SUC.

\subsection{Experimental flow conditions}\label{subsec:Experimental flow conditions}

Table \ref{table:flowcondition} shows the flow conditions used in the present work for the SUC and BCC configurations. The experiments are performed under atmospheric pressure ($\sim 1$ atm) and ambient temperature conditions ($\sim20^{\circ}\mathrm{C}$). The Reynolds number of the particle or sphere, $\mathrm{Re}_\mathrm{p}$, is defined based on the particle diameter and the interstitial velocity ($U_\mathrm{int}$) between particles. The interstitial velocity is defined as the ratio between superficial velocity and the actual volume fraction of the corresponding SUC or BCC packing. The superficial velocity ($U_\mathrm{spf}$) is defined as the average velocity through the square cross-section in the absence of particles. The volumetric flow rates of the co-flow ($Q_\mathrm{C}$) and the central jet flow ($Q_\mathrm{J}$) are kept equal to each other for each flow condition, and are tabulated in standard litres per minute ($\mathrm{lpm}$). The standard condition for the mass flow controllers correspond to 1 atm and 20$^{o}C$. These flow conditions are chosen specifically to provide three different $\mathrm{Re}_\mathrm{p}$: 200, 300, and 400 for each particle bed configuration, which correspond to the laminar and transitional regime for the SUC packing, and the transitional and turbulent for the BCC packing, according to the experimental studies by \cite{Bu2015}. The material properties of the fluid (synthetic air) were determined at $20^{\circ}\mathrm{C}$ and 1 atm (kinematic viscosity $=1.52 \times 10^{-5} \, \mathrm{m}^{2}/\mathrm{s}$ ).

\begin{table}[!t]
	\centering
	\caption{Operating flow conditions for the SUC and BCC packed bed configuration. $\mathrm{Re}_\mathrm{p}$: particle Reynolds number, $U_\mathrm{int}$: interstitial velocity, $U_\mathrm{spf}$: superficial velocity, $Q_\mathrm{C}$: volumetric flow rate of the co-flow, $Q_\mathrm{J}$: volumetric flow rate of the central jet flow, $U_\mathrm{C}$: flow velocity of the co-flow and $U_\mathrm{J}$: flow velocity of the central jet at the exit of the central pipe.}\label{table:flowcondition}
	\centering\begin{tabular}{|c|c|c|c|c|c|c|c|}
		\hline
		\multicolumn {8}{|c|} {SUC configuration } \\
		\hline
		{Case}  & {$\mathrm{Re}_\mathrm{p}$} & {$U_\mathrm{int}$ (m/s)} & {$U_\mathrm{spf}$ (m/s)} & {$Q_\mathrm{C} (\mathrm{l}\mathrm{pm})$} & {$Q_\mathrm{J} (\mathrm{lpm})$} & {$U_\mathrm{C}$ (m/s)} & {$U_\mathrm{J}$ (m/s)}\\
		\hline
		1 & 200& 0.12& 0.06& 38.8& 38.8& 0.028& 12.9\\
		2 & 300& 0.18& 0.08& 58.2& 58.2& 0.042& 19.3\\
		3 & 400& 0.24& 0.11& 77.6& 77.6& 0.056& 25.7\\
  \hline
  	\multicolumn {8}{|c|} {BCC configuration } \\
   \hline
   		4 & 200& 0.12& 0.04& 24.4& 24.4& 0.018& 8.1\\
		5 & 300& 0.18& 0.05& 36.6& 36.6& 0.026& 12.1\\
		6 & 400& 0.24& 0.07& 48.8& 48.8& 0.035& 16.2\\
		\hline
	\end{tabular}
\end{table}

\subsection{Optical setup and methodology for stereo particle imaging velocimetry}\label{subsec:SPIV technique}

\begin{figure}[!t]
	\centering   	
	\includegraphics[width=0.8\linewidth]{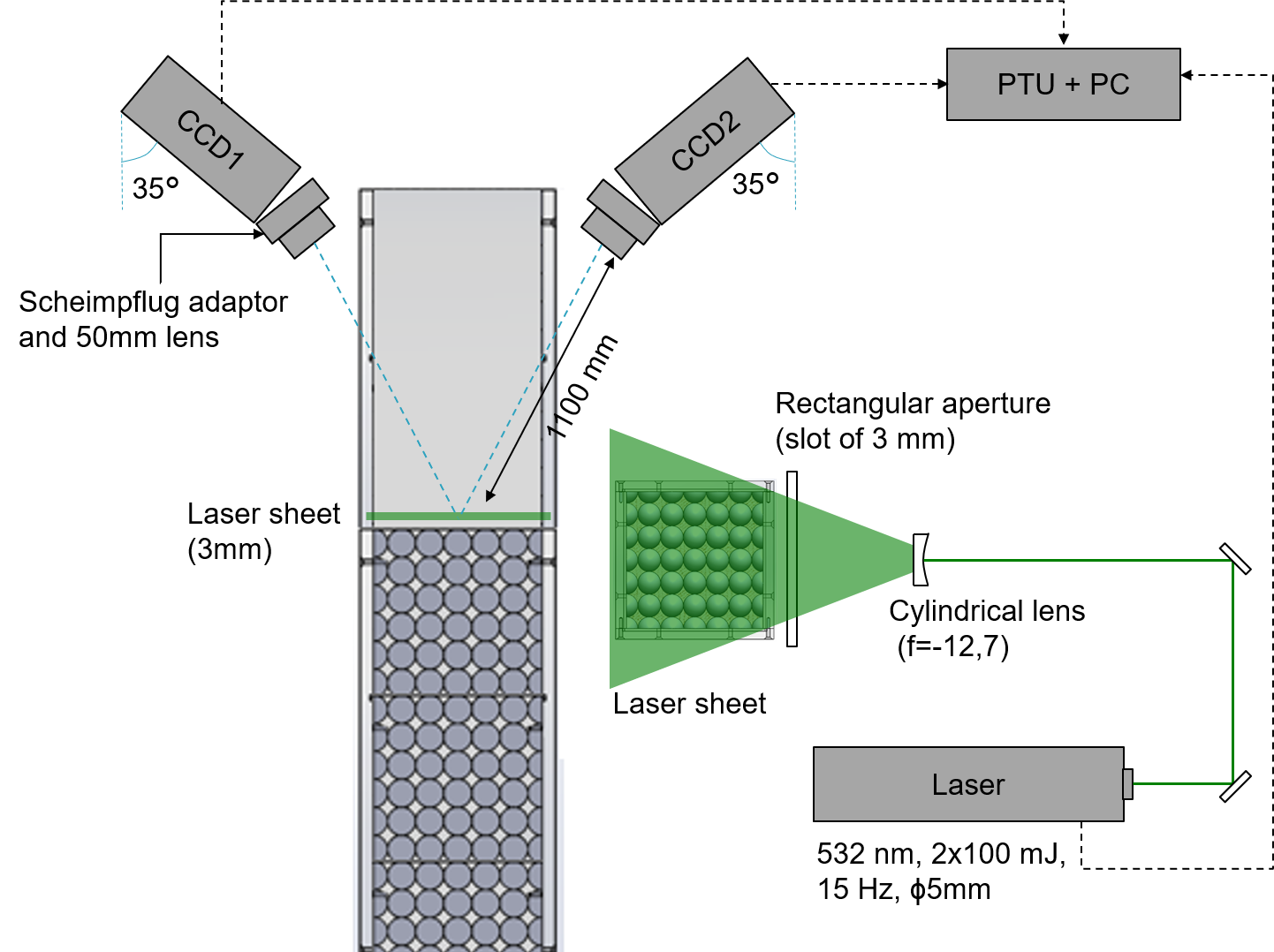}\\
	\caption{The optical setup for stereo particle image velocimetry.}
	\label{fig:SPIVexp}
\end{figure}

Figure \ref{fig:SPIVexp} shows the optical arrangement for the stereo particle imaging velocimetry (SPIV) experiments performed in the present study. The aim of this setup is to determine the velocity vectors of tracer particles over the entire cross-section of the flow at the outlet of the packed bed. It should be noted that the major velocity component of the fluid flow (i.e. along axial $Z$ direction) is perpendicular to the cross-section. A Nd:YAG double pulse laser (Litron Lasers, nano L PIV, 532 nm, 800 mJ, 4 ns pulse duration, double pulse, 15 Hz pulse frequency, and 5mm beam diameter) is used for illuminating the seeding particles. The laser beam is transformed into a sheet using a cylindrical lens (focal length = -12.7 mm). The laser sheet had a thickness of approx. 5 mm, but it is reduced to 3 mm using a rectangular aperture. The aperture consists of a metallic plate with a rectangular slot of 3 mm width and of 152.5 mm length, corresponding to the width of the square channel. The diverging laser sheet is aligned perpendicular to the exit of the packed bed (i.e. axial $Z$ direction). The mid-plane of the laser sheet is placed 5.5 mm above the exit of the packed bed. Double-frame images were captured simultaneously using two CCD cameras (Lavision, Imager Pro X, 1600$\times$1200 resolution and 7.4$\times$7.4 $\mu$ m$^2$ pixel size). Scheimpflug adaptors from Lavision and 50 mm objective lenses from Nikon have been mounted in each camera to clearly focus all the seeding particles at the measurement plane. Both the cameras are arranged at an inclination of 35$^{\circ}$ with respect to the $Z$ axis. The apertures of both camera lenses are closed to $f$-numbers of 11. This ensures that the depth of field is larger than the laser sheet thickness. Cameras and laser were synchronized and controlled by a programmable time unit (PTU from LaVision) and Davis 8.4 software (Lavision). In total, for each flow condition, 1000 pairs of double-frame images have been recorded by each camera. The laser pulse delay is varied between 400 $-$ 1400 $\mu$s, depending on the $\mathrm{Re}_\mathrm{p}$ of the flow conditions, to prevent the particles from leaving the laser sheet in between recording of the image frames. The recording of the stereo image pairs is carried out by Davis 8.4. 

Regarding the camera calibration, a calibration plate is manufactured based on a dot pattern generated by Davis 8.4, where the diameter of each dot is 3 mm and the distance between dots is 7 mm. The pattern was printed and glued to a flat and rigid metal plate. During calibration, the calibration plate is kept inside the square channel, close to the exit of the packed bed and parallel to the plane of the laser sheet. Then, 7 positions, every 0.5 mm, are traversed, with an accuracy of 5 $\mu$m, over a range of 3 mm, within the region of interest. Thereafter, the calibration is performed by a mapping function based on a polynomial fit of 3$^\mathrm{rd}$ order~\citep{Soloff1997}. To correct any laser sheet misalignment, the self-calibration procedure by \cite{wieneke2005selfclb} is performed. Finally, the RMS error of the calibration is determined to be 0.05 pixels (7.9 $\mu$m in the object plane).

The SPIV images are captured near the packed bed and, since the spheres are printed in white colour and their surfaces reflect the laser light, the spheres are always visible in the recorded images. In order to minimize the noise arising from this scattering, the last layer of the packed bed of SUC and BCC is painted in matt black. Furthermore, a set of 100 background images are captured in the presence of the laser sheet without any seeding particles and averaged to have one average background image. 

The image pre-processing is carried out using Davis 8.4 software \citep{LaVision2017}. First, the average background image is subtracted from each instantaneous SPIV image to remove any offset in the instantaneous images. It is observed that even after background subtraction, some of the sphere surfaces were still visible, because there is some additional scattering from the seeding particles. To reduce the effect of such reflections, Davis 8.4 offers a sliding background subtraction and particle intensity normalization pre-processing operations to minimize local intensity fluctuation in the background and to correct local particle intensity fluctuation, respectively. Those tools are applied using a local scale length of 7 pixels. 

Regarding the vector calculations, a multi-pass approach is adopted. First, two iterations were performed with an interrogation window with a size of 64$\times$64 pixels (75\% overlap) and the remaining four passes had an interrogation window size of 16$\times$16 pixels (50 \% overlap), which lead to a spatial resolution of 8 pixels (1.24 mm in the object plane). The uncertainty in the instantaneous velocity components is defined by the correlation statistics method by \cite{wieneke2015piv}, while the uncertainty propagated to the mean velocity field and other statistical parameters are evaluated using the method by \cite{sciacchitano2016piv}. Thus, the uncertainty in mean axial velocity is 0.5 - 2 \% while the uncertainty in standard deviation of axial velocity is around 2 \% over the entire plane of measurement. Additionally, Table \ref{table:SPIV Parameters} describes all the important parameters used to collect and process the SPIV images.    

\begin{table}[!t]
	\centering
	\caption{Parameters for the experimental SPIV technique.}\label{table:SPIV Parameters}
	\centering\begin{tabular}{|l|l|}
		\hline
		{Parameter}  & {Value} \\
		\hline
            Maximum out of plane velocity & $\approx 1.5$ m/s\\
		Maximum in-plane velocity & $\approx 0.3$ m/s \\ 
		Measurement area size & $152.5$ mm $\times$ $152.5$ mm\\
		Observation distance & $1100$ mm\\
            Pixel size in the sensor & $7.4$ $\mu$m\\
            Magnification & $0.0472$\\
            $f$-number & $11$\\
            Particle image size & $\approx 2$ pixels\\
            Seeding particle diameter & less than $1 \, \mu$m\\
  \hline
  		\end{tabular}
\end{table}

Table \ref{table:VolumeComp} shows the volumetric flow rate calculated using the SPIV measurements for the SUC and BCC configurations and the respective percentage difference with respect to the actual volumetric flow rate for all $\mathrm{Re}_\mathrm{p}$. It is observed that for the SUC configuration, the difference in volumetric flow rate is 9-14 \% while for BCC, it is 15-20 \%. Additionally, with increase in $\mathrm{Re}_\mathrm{p}$, the difference is shown to increase for SUC and BCC both. This suggests that, when the flow is close to laminar conditions, for example in SUC for $\mathrm{Re}_\mathrm{p}$ = 200, the difference in volumetric flow rate is up to 10 \%. On the other hand, when turbulence is induced in the flow, and higher fluctuations in velocity arise, the difference in volumetric flow rate goes up to 20 \%.

\begin{table}[!t]
	\centering
	\caption{Comparison of volumetric flow rates between the actual conditions and the values obtained from the SPIV measurements.}\label{table:VolumeComp}
	\centering\begin{tabular}{|c|c|c|c|c|}
		\hline
		{Case}  & {$\mathrm{Re}_\mathrm{p}$} & {Actual volume} & {volumetric flow rate} & {percentage} \\
  		  & & {flow rate ($\mathrm{lpm}$)} & {from SPIV technique ($\mathrm{lpm}$)} & {difference}\\
		\hline
  	\multicolumn {5}{|c|} {SUC configuration } \\
   
		\hline
		1 & 200& 77.6 & 70.3& 9.4 \%\\
		2 & 300& 116.4& 105.7& 9.2 \%\\
		3 & 400& 155.2& 133.2& 14.1 \%\\
  \hline
  	\multicolumn {5}{|c|} {BCC configuration } \\
   \hline
   		4 & 200& 48.8& 41.5& 14.9 \%\\
		5 & 300& 73.2& 60.2& 17.7 \%\\
		6 & 400& 97.7& 77.2& 20.9 \%\\
		\hline
	\end{tabular}
\end{table}

\section{Numerical method}\label{sec:Numerical method}
The gaseous flow inside the fixed packed bed reactor is considered as an incompressible Newtonian fluid with constant fluid properties and is therefore governed by the Navier-Stokes equations,
\begin{align}
    \nabla \cdot \mathbf{u} &= 0 \, ,\\
    \rho \left( \frac{\partial \mathbf{u}}{\partial t} + \nabla \cdot ( \mathbf{u} \otimes \mathbf{u}) \right) &= - \nabla p + \nabla \cdot \boldsymbol{\tau} + \rho \mathbf{g} + \mathbf{s}\,,
\end{align}
which are discretized and solved on an Eulerian mesh with adaptive mesh refinement (AMR). $\rho$ is the fluid density, $\mathbf{u}$ the velocity vector, $p$ the pressure, $\boldsymbol{\tau}$ the viscous stress tensor, $\mathbf{g}$ the gravity acceleration vector, and $\mathbf{s}$ represents a momentum source term arising from the presence of the immersed boundaries. The Navier-Stokes equations are discretized and solved using a finite-volume framework with a collocated variable arrangement in a coupled, pressure-based manner with second-order accuracy in space and time \citep{Denner2014a, Denner2020}.

The particle surface is discretized by uniformly distributed Lagrangian markers $\mathbf{X}_\mathrm{J} \,, j \in \{ 1,...,N_L \}$, with an optimal distance between the markers of the order of the Eulerian fluid mesh spacing \citep{Zhou2021}. For the computation of the Lagrangian feedback force, the momentum equation is reformulated as
\begin{equation}
    \mathbf{F}_\mathrm{J}^n = \frac{\rho}{\Delta t} \left( \mathbf{U}_{\mathrm{IB},j}^n - \mathbf{U}_\mathrm{J}^{n-1} \right) + \mathbf{C}_\mathrm{J}^n + \mathbf{B}_\mathrm{J}^n - \mathbf{D}_\mathrm{J}^n - \rho \mathbf{g} \,,
\end{equation}
for each Lagrangian marker $j$, the direct forcing approach by \citet{AbdolAzis2019} is applied. The super-script $n$ denotes the time level at which the quantities are to be evaluated, $\mathbf{U}_{\mathrm{IB},j}$ is the velocity vector of the \textit{j}-th Lagrangian marker, and $\mathbf{U}_\mathrm{J}$, $\mathbf{C}_\mathrm{J}$, $\mathbf{B}_\mathrm{J}$, and $\mathbf{D}_\mathrm{J}$, are the interpolated Eulerian velocity, advection, pressure, and diffusion terms of the governing momentum equations, respectively. The previous equation can be further simplified and the momentum terms can be further summarized as
\begin{equation}
    \mathbf{F}_\mathrm{J}^n = \frac{\rho}{\Delta t} \left( \mathbf{U}_{\mathrm{IB},j}^n - \mathbf{\hat{U}}_\mathrm{J} \right) + \mathbf{\hat{F}}_\mathrm{J}^n \,.
\label{Eq:Lagrangian force}
\end{equation}

For the coupling of the Lagrangian forces with the momentum equation, the deferred fluid velocity $\mathbf{\hat{U}}_\mathrm{J}$ and the deferred momentum terms cumulated in $\mathbf{\hat{F}}_\mathrm{J}^n$ of time level $n$ are interpolated from the Eulerian mesh 
by an adequate interpolation operator. Such a discrete, compact interpolation operator interpolates the fluid velocities within a symmetric stencil with a certain radius (usually a few fluid cell spacings) to the Lagrangian markers and the interpolated velocities are therefore referred to as the Lagrangian velocity \citep{Peskin1972}. The interpolation of an arbitrary fluid variable $\gamma$ to the position of a Lagrangian marker $j$ reads as
\begin{equation}
    \Gamma_\mathrm{J} = \sum_{i \in \delta_\mathrm{J}} \phi_{i,j} \gamma_i \,, 
\end{equation}
where $\Gamma_\mathrm{J}$ is the fluid variable interpolated to the position of Lagrangian marker $j$, $\delta_\mathrm{J}$ is the set of
Eulerian cells in the interpolation support of the $j$-th Lagrangian marker, $\phi_{i,j}$ is the discrete interpolation
weight associated with the $i$-th Eulerian cell in the support stencil, and $\gamma_i$ is the fluid variable for that Eulerian
cell $i$. The discrete interpolation weight $\phi_{i,j}$ is based on a normalized kernel function $\phi \, : \, \mathbb{R}^3 \rightarrow \mathbb{R}$ and can be calculated as 
\begin{equation}
   \phi_{i,j} = \phi \left( \mathbf{x}_i - \mathbf{X}_\mathrm{J} \right) V_i \,,
   \label{Eq: Interpolation weight}
\end{equation}
where $\mathbf{x}_i$ is the centroid of the $i$-th Eulerian fluid cell in the support stencil, $\mathbf{X}_\mathrm{J}$ is the position of the $j$-th Lagrangian marker and $V_i$ is the volume of the fluid cell. 
Throughout this work, a support stencil with a radius of four times the fluid mesh spacing is used. This interpolation stencil numerically thickens the particle-fluid interface over a few fluid cells across the particle surface. Henceforth, the required Lagrangian force to satisfy the no-slip condition at the particle surface is calculated from the difference between the interpolated Lagrangian velocity and the desired velocity at the surface. The desired velocity typically arises from the rigid body motion of the solid object \citep{Uhlmann2005}.

After the interpolation step, the Lagrangian forces are computed at the location of each Lagrangian marker, see Eq. \eqref{Eq:Lagrangian force}, followed by spreading the Lagrangian forces, with the same stencil as used for the interpolation, back to the fluid cells in the support region. On the fluid mesh, the force is applied as a volumetric source term in the discretized equations governing the fluid flow. Depending on the implicitness of the above procedure, the procedure of interpolation and force computation may require a number of iterations before the accurate no-slip boundary condition at the particle surface is obtained. The spreading of an arbitrary Lagrangian variable $\Gamma$ onto the Eulerian mesh follows as
\begin{equation}
    \gamma_i = \sum_{j \in \psi_i} \phi_{i,j} W_\mathrm{J} \Gamma_\mathrm{J} \,,
\end{equation}
where $W_\mathrm{J}$ is the spreading weight associated with the $j$-th Lagrangian marker, $\psi_i$ is the set of Lagrangian markers whose spreading support stencil contains the $i$-th Eulerian cell, and $\phi_{i,j}$ is the same as in Eq. \eqref{Eq: Interpolation weight}. The spreading weight $W_\mathrm{J}$ is a non-physical quantity, and many definitions have been used and discussed in the literature \citep{Uhlmann2005, AbdolAzis2019, Pinelli2010, Zhou2019}. 
For an optimal compromise between stability and accuracy of the no-slip boundary condition enforcement at the particle surface, the spreading weights for the IBM in this work are treated with a stability analysis  \citep{Zhou2021,Cheron2023a}.

A qualitative illustration of the interpolation and spreading stencil in 2D for the Lagrangian markers is also given in Figure \ref{fig:NumericalPacking}. For the three white coloured Lagrangian markers, the support stencil is shown by circles. The stencils are symmetric and the Eulerian fluid mesh cells within these stencils contribute to the interpolation to the corresponding markers and are the ones to which the Lagrangian forces are spread onto.
A more detailed step by step derivation and implementation of the IBM and its interpolation and spreading is given, for instance, in \citet{Cheron2023a}.

\section{Numerical setup} \label{sec:Numerical setup}

\begin{figure}[h!]
	\centering
	\subcaptionbox{Numerical domain\label{fig:NumericalDomainSize}}{\includegraphics[width=4cm]{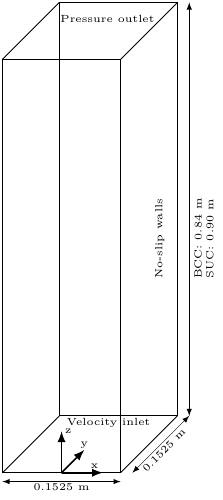}}
	\subcaptionbox{Plane in the region of interest and qualitative illustration of the Lagrangian discretization over an Eulerian fluid mesh\label{fig:NumericalPacking}}{\includegraphics[width=8cm,clip = 0cm 0cm 0cm 0.0cm]{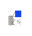}}
    \subcaptionbox{Exemplary mesh cutout\label{fig:AMRMesh}}{\includegraphics[width=3cm]{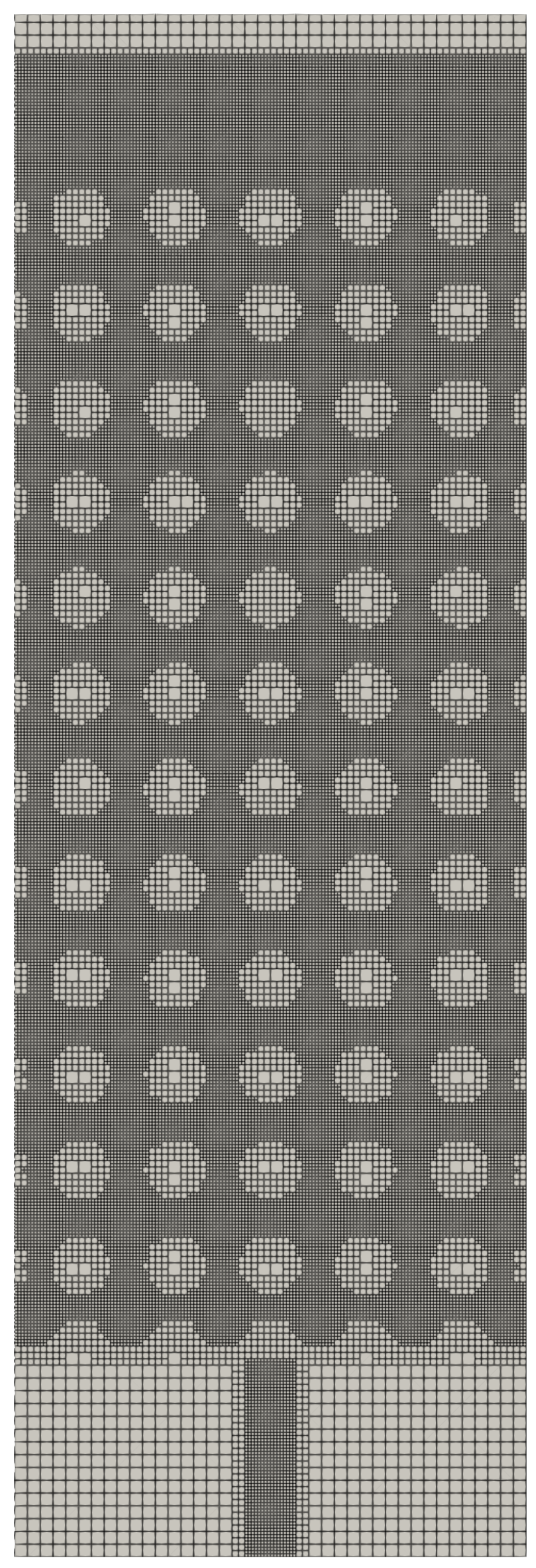}}
    \caption{General numerical setup of the packed particle bed reactor with: (a) the numerical domain, (b) an example of the measuring plane in the region of interest above the packing and a qualitative illustration of the Lagrangian discretization over an Eulerian fluid mesh for the IBM, and (c) an example of the cross-section of the adaptive refined mesh.}
	\label{fig:NumericalSetup}
\end{figure}

The general numerical setup is adopted from the experimental configuration introduced in section \ref{sec:Experimental setup and methodology}. Based on the experimental conditions of atmospheric pressure and ambient temperature conditions of $\sim20^{\circ}\mathrm{C}$, the fluid properties for air are chosen to be $1.204 \, \mathrm{kg/m^3}$ for the density and $1.82 \cdot 10^{-5} \, \mathrm{kg/{m \, s}}$ for the dynamic viscosity. As can be seen in Figure \ref{fig:NumericalDomainSize}, the overall numerical domain size of the fixed packed bed reactor is $[0.1525 \times 0.1525 \times 0.84] \, \mathrm{m^3}$ for the BCC particle configuration, and $[0.1525 \times 0.1525 \times 0.90] \, \mathrm{m^3}$ for the SUC particle configuration. Both reactors have a velocity inlet (Dirichlet boundary condition for velocity, Neumann boundary condition for pressure) at the bottom, a pressure outlet (Neumann boundary condition for velocity, Dirichlet boundary condition for pressure) at the top and no-slip walls (Dirichlet boundary condition for velocity, Neumann boundary condition for pressure) at the sides as domain boundary conditions.

Compared to the experimental setup, the numerical domain starts directly above the porous plate with the velocity inlet and ends at the outlet of the extended walls with the pressure outlet. The modelling of the velocity inlet is divided into two sections, the centre jet region and the co-flow region.
A cell face at the inlet corresponds to the centre jet region if its area  intersects with the $8 \, \mathrm{mm}$ jet hole radius from the exact centre of the inlet plane.  All other inlet cell faces are considered as co-flow inlet faces. Therefore, the inlet boundary condition for the velocity is modelled as
\begin{equation}
    u_{\text{inlet}} = A_{\text{fraction}} \cdot u_{\text{J}} + (1 - A_{\text{fraction}}) \cdot  u_{\text{C}}\,,
\end{equation}
where $A_{\text{fraction}} = A_{\text{inside}} / A_{\text{cell}}$ is the partial area of the inlet cell face which falls inside the centre jet area,  $A_{\text{cell}}$ is the total area of the cell face, $u_{\text{J}}$ and $u_{\text{C}}$ are jet and co-flow inlet velocities, given in Table \ref{table:flowcondition}. Since all variables are known a priori, this leads to an exact inlet mass flow compared to the dictated mass flow from experimental measurements.

It should be noted here, that the inlet conditions for the numerical setup are precisely symmetrical, with no variations in space and time, which is not possible to be completely ensured for the experimental setup. Furthermore, for the numerical setup, the jet inlet pipe has no predefined wall thickness. It is given a slighlty artificial wall thickness based on $A_{\text{fraction}}$ for the cell faces in the transitional area between the jet region and the co-flow region, which may not perfectly match the pipe thickness as used in the experiment. Furthermore, the velocity profile of the jet is prescribed uniformly and not parabolic.\\
The SUC and BCC particle packing configurations are modelled according to the information of the 3D-printed packings from the experimental setup, although the surfaces of the particles are assumed smooth. For the SUC packing, the overlap between particles of adjacent layers is $0.1 \, \mathrm{mm}$ and the overlap between particles of adjacent layers for the BCC packing is $0.88 \, \mathrm{mm}$. Therefore, the discrepancy between the heights, respectively the volume fractions, of the experimental and numerical packings is below 1 \%.
Here it is noted that the step on which the packings rest in the experimental setup is not modelled at all in the numerical setup. 

An example of the adapted and refined Eulerian fluid mesh for the numerical IBM simulations is shown in Figure \ref{fig:AMRMesh}. Within the interpolation/spreading stencil around the particle surfaces, in the region of interest above the fixed packed bed, and around the inlet jet, the cell width is set to satisfy the desired particle diameter to fluid cell edge length ($d_{\text{p}} / \Delta x$) ratio, which is set to 26 in this work. This is in accordance with our earlier findings~\citep{Gorges2024}. In the remaining interstices and at the outlet, the cell width is double the size. For all simulations, the Courant-Friedrichs-Lewy number ($\mathrm{CFL}$) is fixed in the range of 0.25 to 0.30. After an initial flow phase for the formation of the flow structures, and to ensure independence of the starting conditions, the velocity field is recorded in intervals with a frequency of $20 \, \mathrm{Hz}$ with  a total period of approximately $2 \, \mathrm{s}$ of physical time. The results which are compared are then the average values of all previously stored velocity fields.

The region of interest for the comparisons is a $3 \, \mathrm{mm}$ high volume at a height of $4 - 7 \, \mathrm{mm}$ above the respective particle packing. This region of interest is then divided into a grid with the approximate cell size of $[1.25 \times 1.25 \times 3.00] \, \mathrm{mm}$ for a more accurate comparison with the experimental data. All time-averaged velocity data within one of these post-processing grid cells are then averaged to a single post-processing grid cell value.

\section{Results and discussion} \label{sec:Result&Discussion}

In this section, the experimental and numerical results are compared and discussed. Velocity contour and line plots and probability distributions of averaged axial velocity are presented and discussed for the SUC and BCC configurations for $\mathrm{Re}_\mathrm{p}$ = 200, 300 and 400. 

\subsection{Flow characteristics for the SUC packed bed} \label{subsec:RD_SUC}

\begin{figure}[h!]
	\centering
	\subcaptionbox{Simulation $\mathrm{Re}_\mathrm{p}$ = 200\label{fig:AvgVzSUC200Sim}}{\includegraphics[width=8cm]{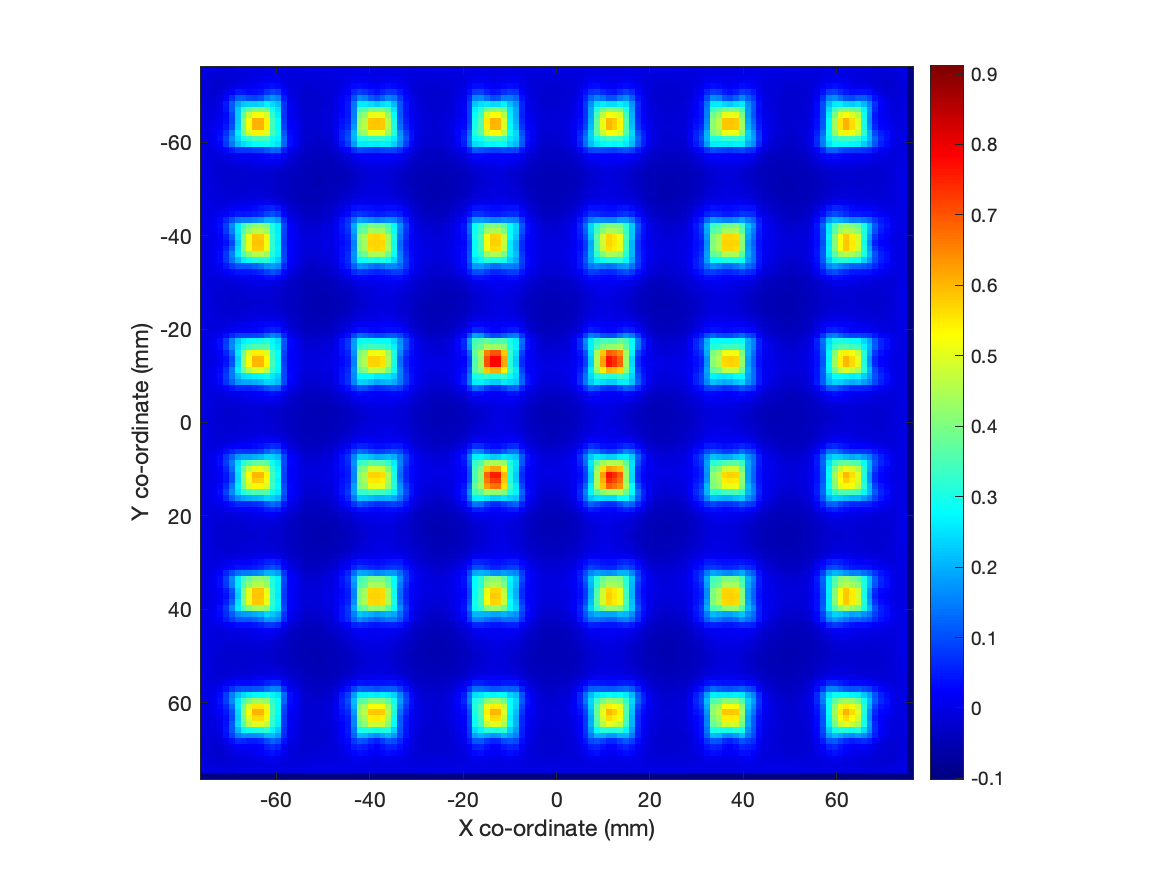}}
	\subcaptionbox{Experiment $\mathrm{Re}_\mathrm{p}$ = 200\label{fig:AvgVzSUC200Exp}}{\includegraphics[width=8cm,clip = 0cm 0cm 0cm 0.0cm]{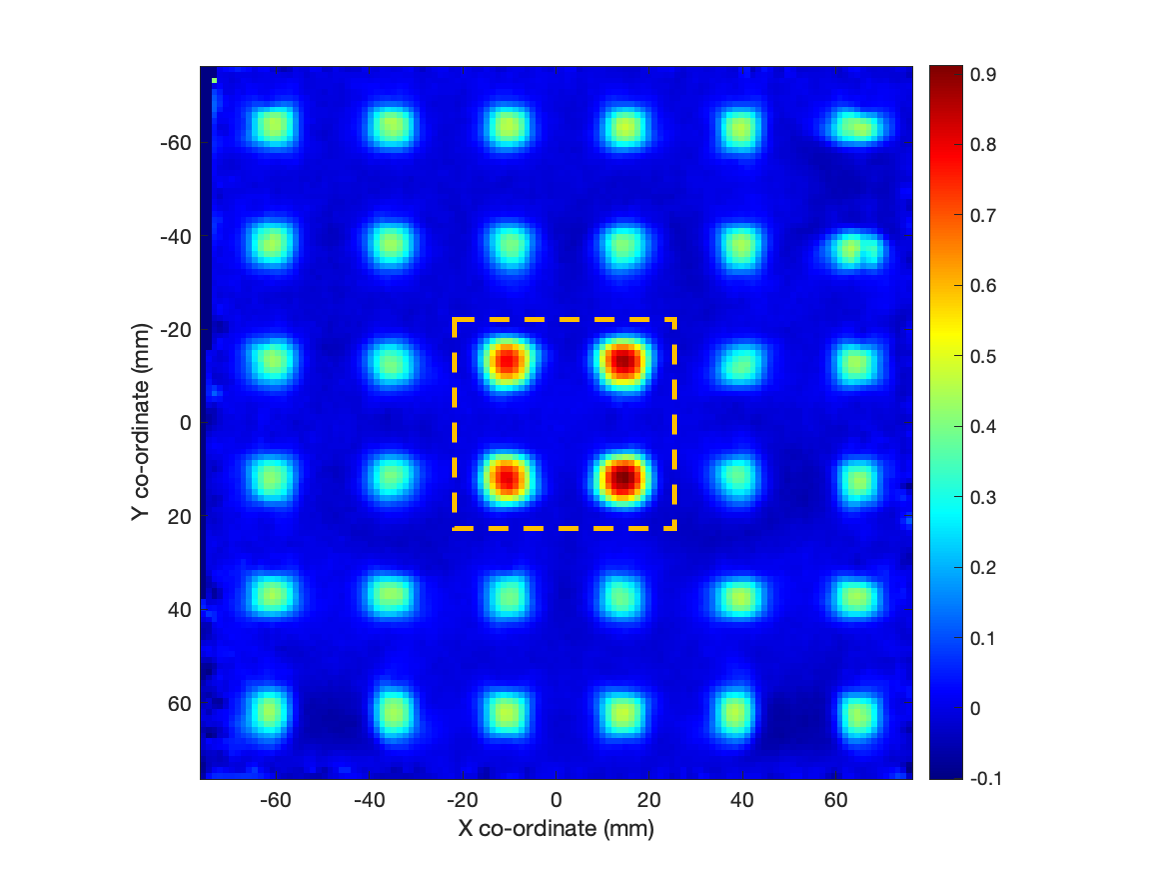}}\\
    \subcaptionbox{Simulation $\mathrm{Re}_\mathrm{p}$ = 300\label{fig:AvgVzSUC300Sim}}{\includegraphics[width=8cm]{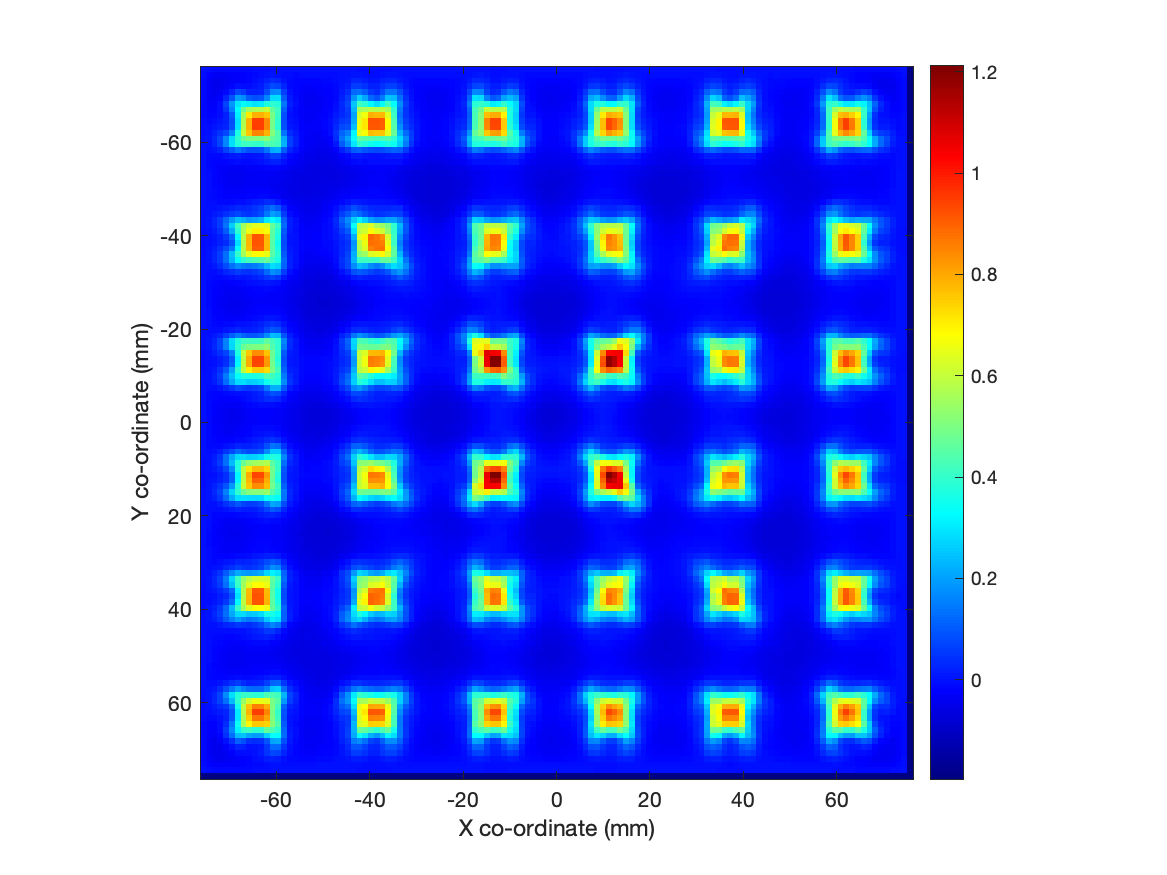}}
	\subcaptionbox{Experiment $\mathrm{Re}_\mathrm{p}$ = 300\label{fig:AvgVzSUC300Exp}}{\includegraphics[width=8cm,clip = 0cm 0cm 0cm 0.0cm]{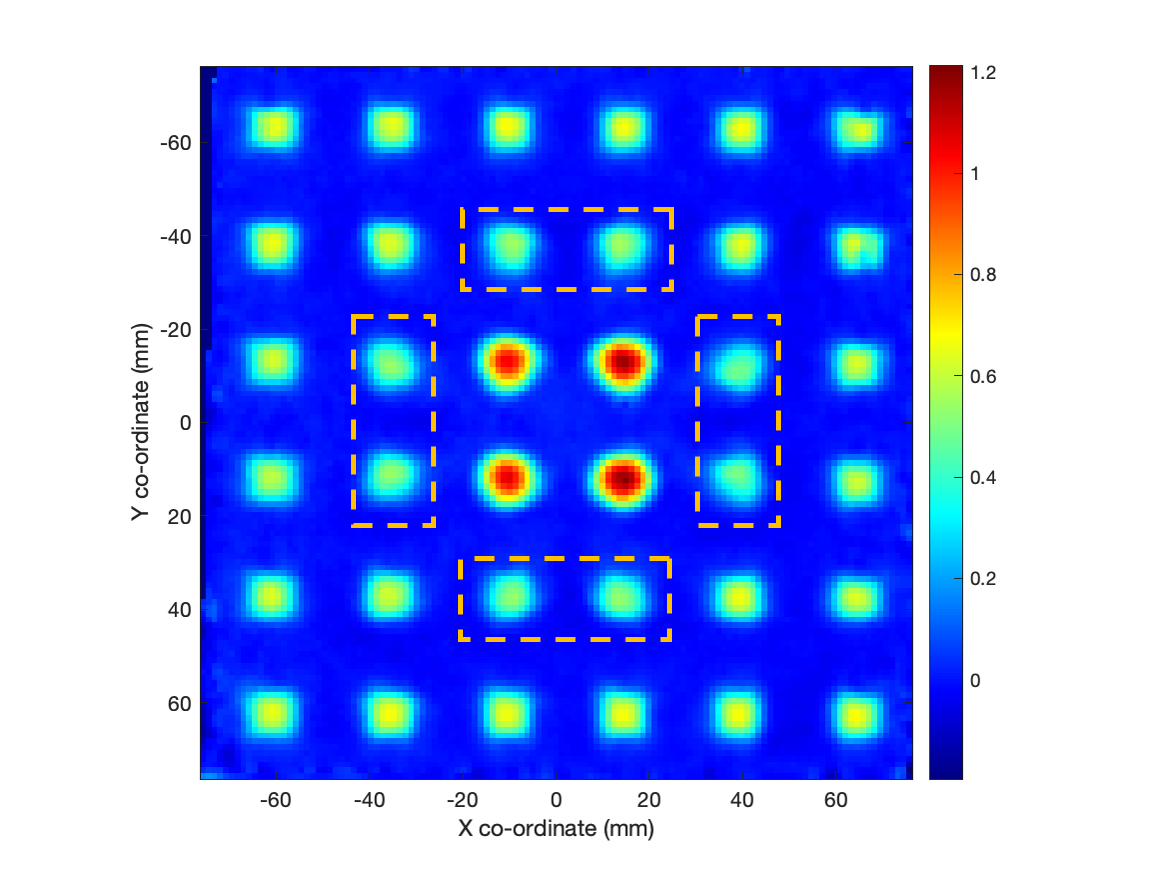}}\\
    \subcaptionbox{Simulation $\mathrm{Re}_\mathrm{p}$ = 400\label{fig:AvgVzSUC400Sim}}{\includegraphics[width=8cm]{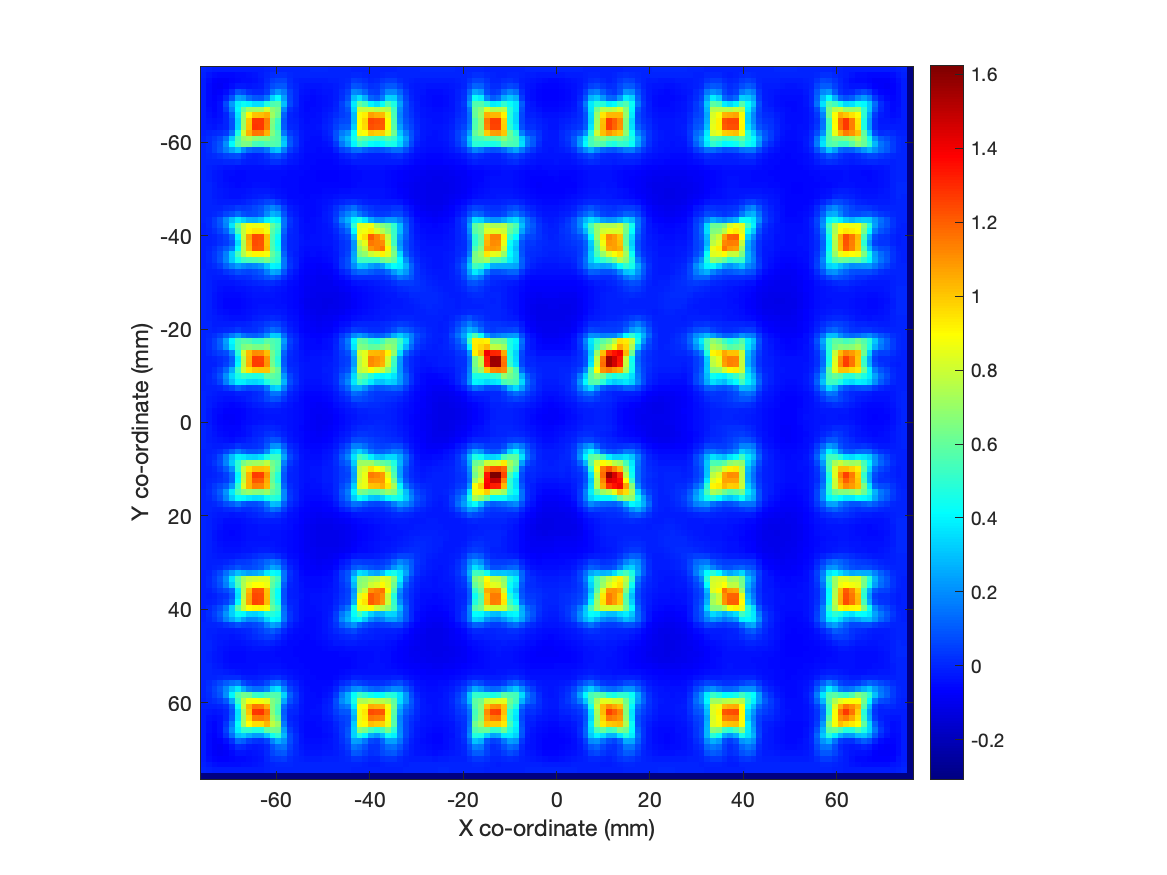}}
	\subcaptionbox{Experiment $\mathrm{Re}_\mathrm{p}$ = 400\label{fig:AvgVzSUC400Exp}}{\includegraphics[width=8cm,clip = 0cm 0cm 0cm 0.0cm]{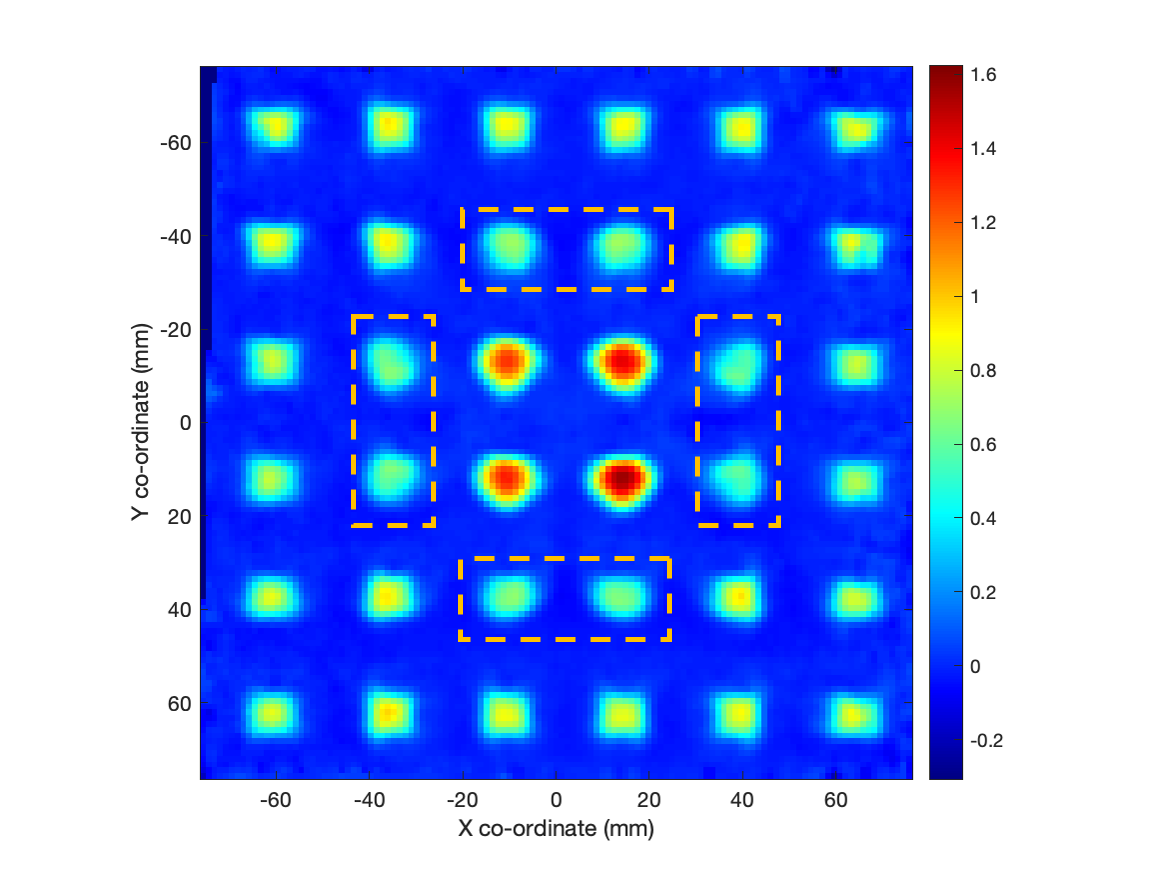}}\\
	\caption{Comparison of numerical and experimental contour plots for averaged axial velocity ($\overline{V_z}$) for SUC packing at (a,b) $\mathrm{Re}_\mathrm{p}$ = 200, (c,d) $\mathrm{Re}_\mathrm{p}$ = 300, and (e,f) $\mathrm{Re}_\mathrm{p}$ = 400.}
	\label{fig:AvgVzSUCComp}
\end{figure}

Figure \ref{fig:AvgVzSUCComp} presents the experimental and numerical contour plots of the averaged axial velocities for the SUC packed bed at three flow conditions, characterized by $\mathrm{Re}_\mathrm{p}$ = 200, 300, and 400. For the SUC configuration, there are 36 pores (see Figure \ref{fig:SUCa-structure}) in between the spheres. Through each pore, the airflow emerges in the form of a small jet. It is noted that in this research work, the jet flow through the central pipe in the experimental setup is referred to as `central jet', while the flow through each pore is referred to as `pore jet' or 'jet'. It can be observed from Figure \ref{fig:AvgVzSUCComp}, that the central jet flow is able to disperse in both lateral directions, and it reaches the pores near the periphery of the packed bed.

\begin{figure}[h!]
	\centering
	\subcaptionbox{Simulation \label{fig:ContourLinesSimSUC}}{\includegraphics[width=5cm]{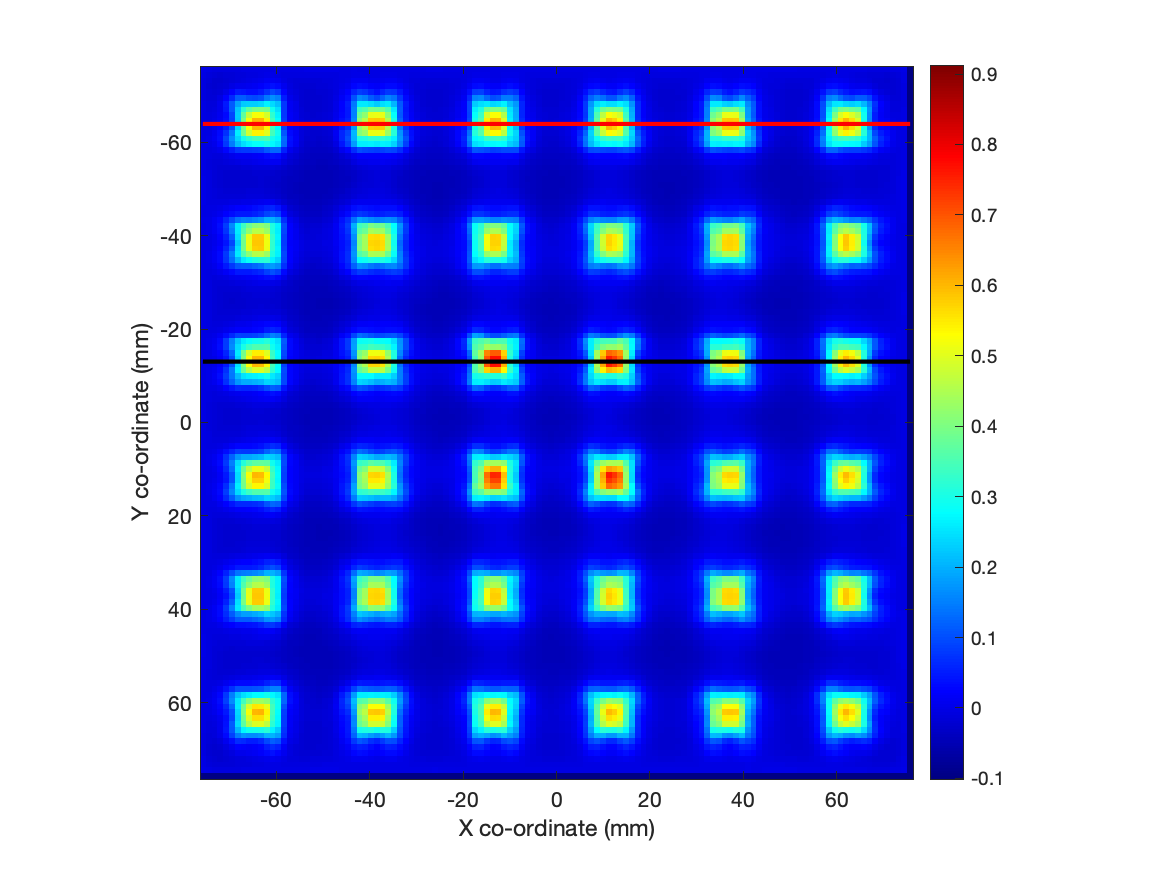}}
	\subcaptionbox{Experiment \label{fig:ContourLinesExpSUC}}{\includegraphics[width=5cm,clip = 0cm 0cm 0cm 0.0cm]{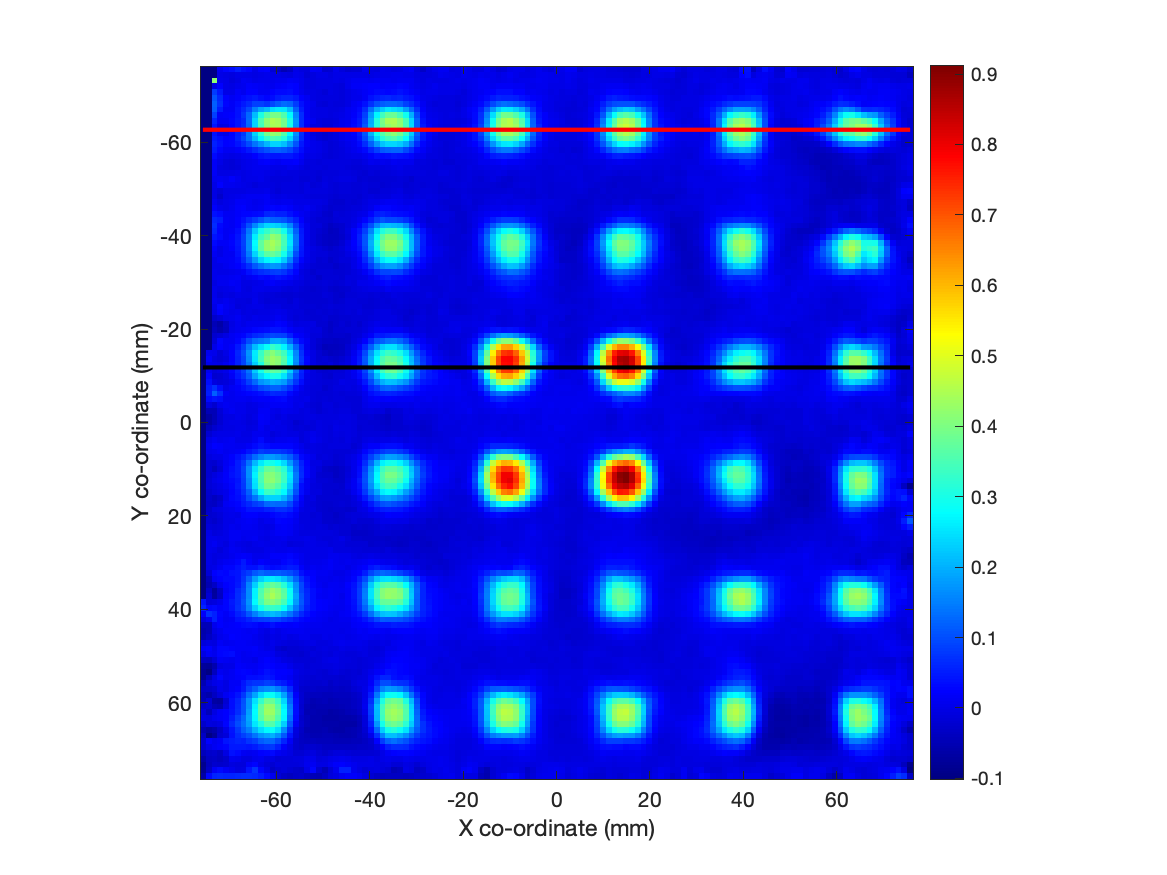}}\\
	\subcaptionbox{Red line $\mathrm{Re}_\mathrm{p}$ = 200\label{fig:Lineplot_red_Rep200SUC}}{\includegraphics[width=6cm]{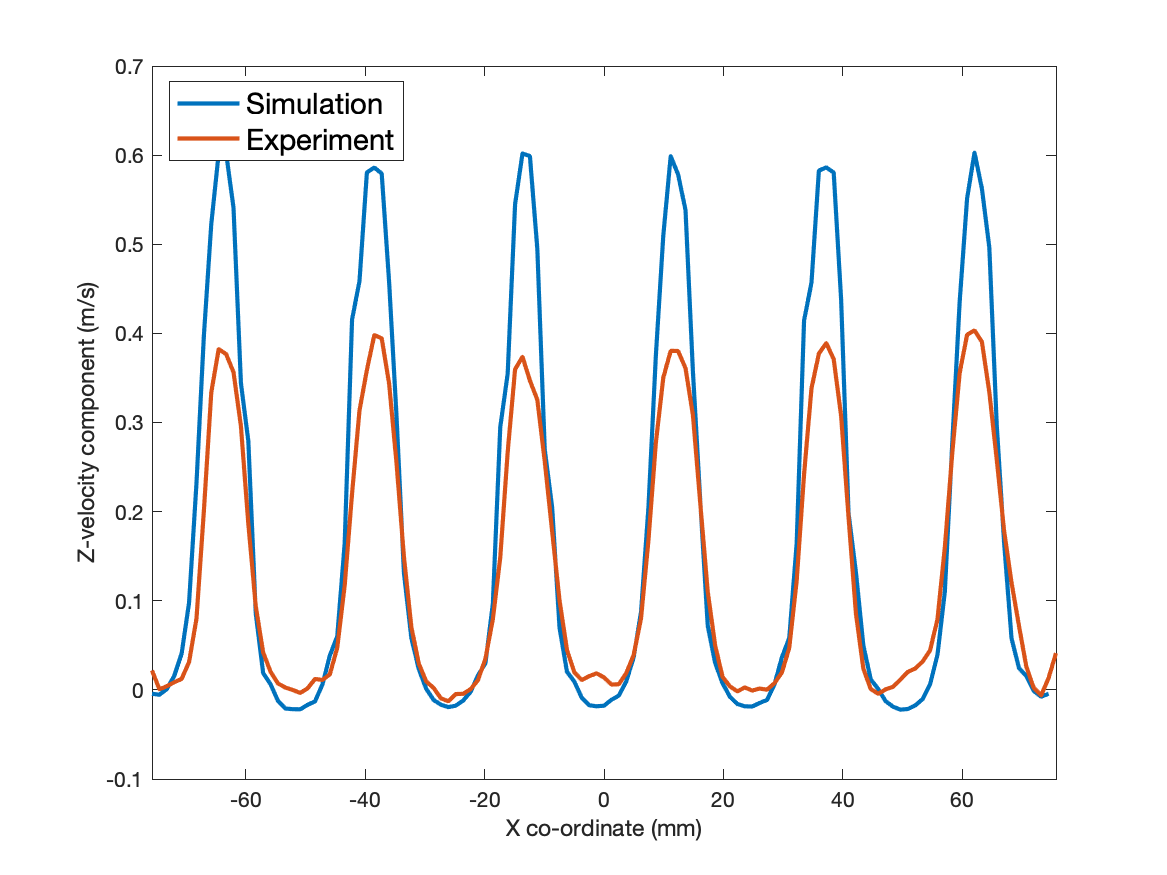}}
	\subcaptionbox{Black line $\mathrm{Re}_\mathrm{p}$ = 200\label{fig:Lineplot_black_Rep200SUC}}{\includegraphics[width=6cm,clip = 0cm 0cm 0cm 0.0cm]{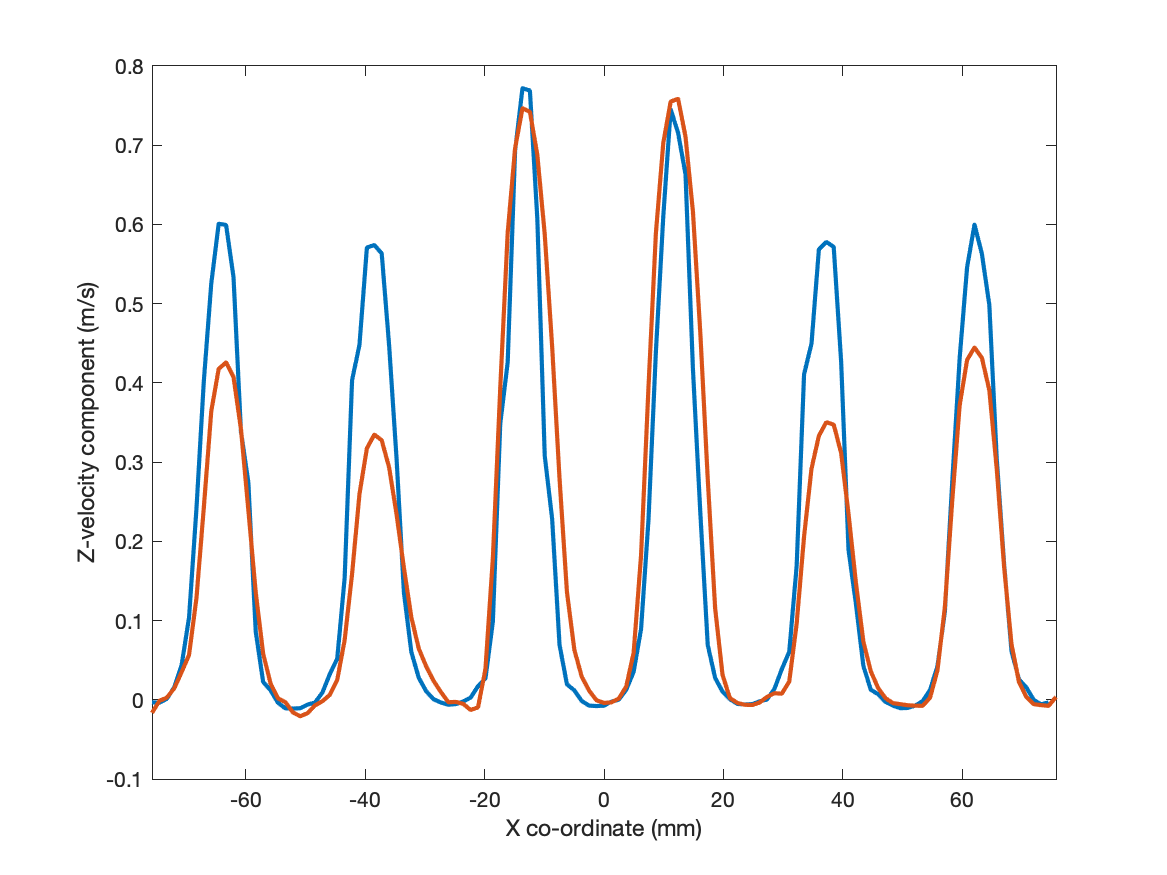}}\\
    \subcaptionbox{Red line $\mathrm{Re}_\mathrm{p}$ = 300\label{fig:Lineplot_red_Rep300SUC}}{\includegraphics[width=6cm]{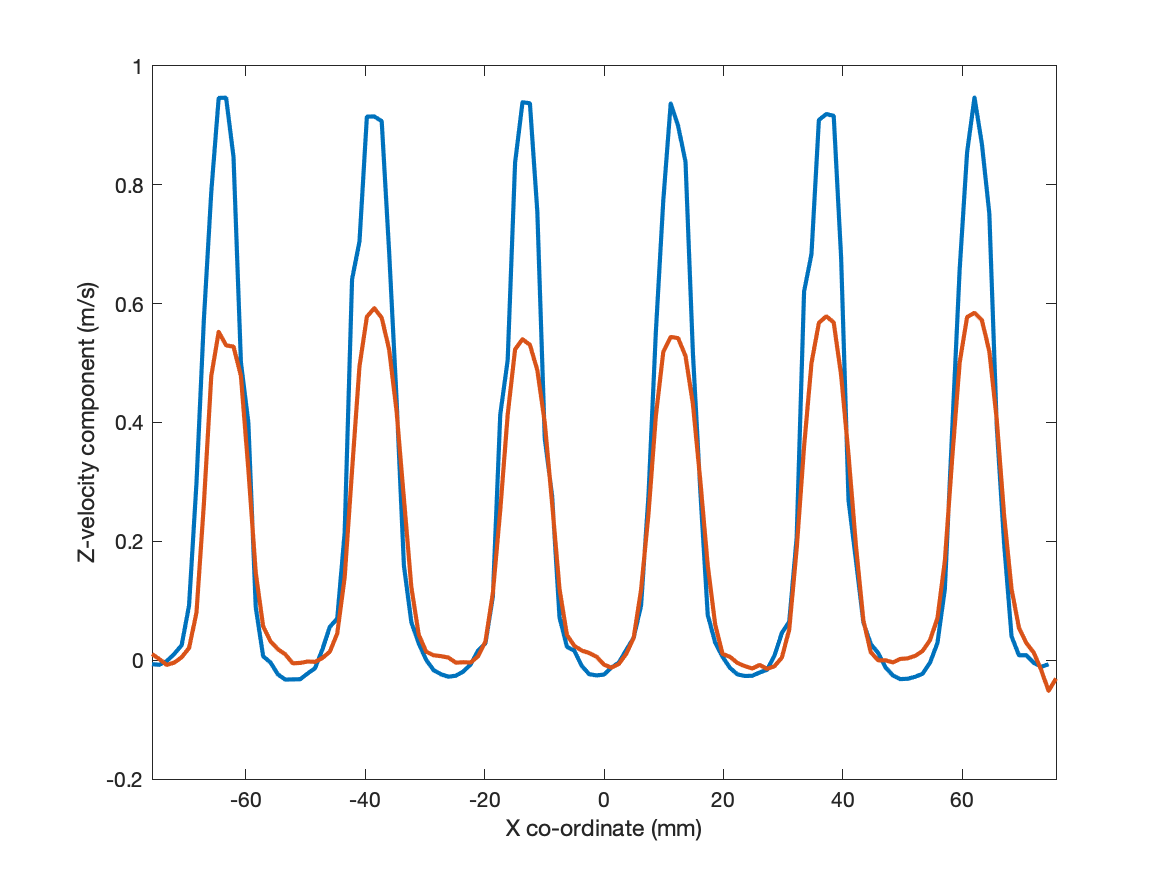}}
	\subcaptionbox{Black line $\mathrm{Re}_\mathrm{p}$ = 300\label{fig:Lineplot_black_Rep300SUC}}{\includegraphics[width=6cm,clip = 0cm 0cm 0cm 0.0cm]{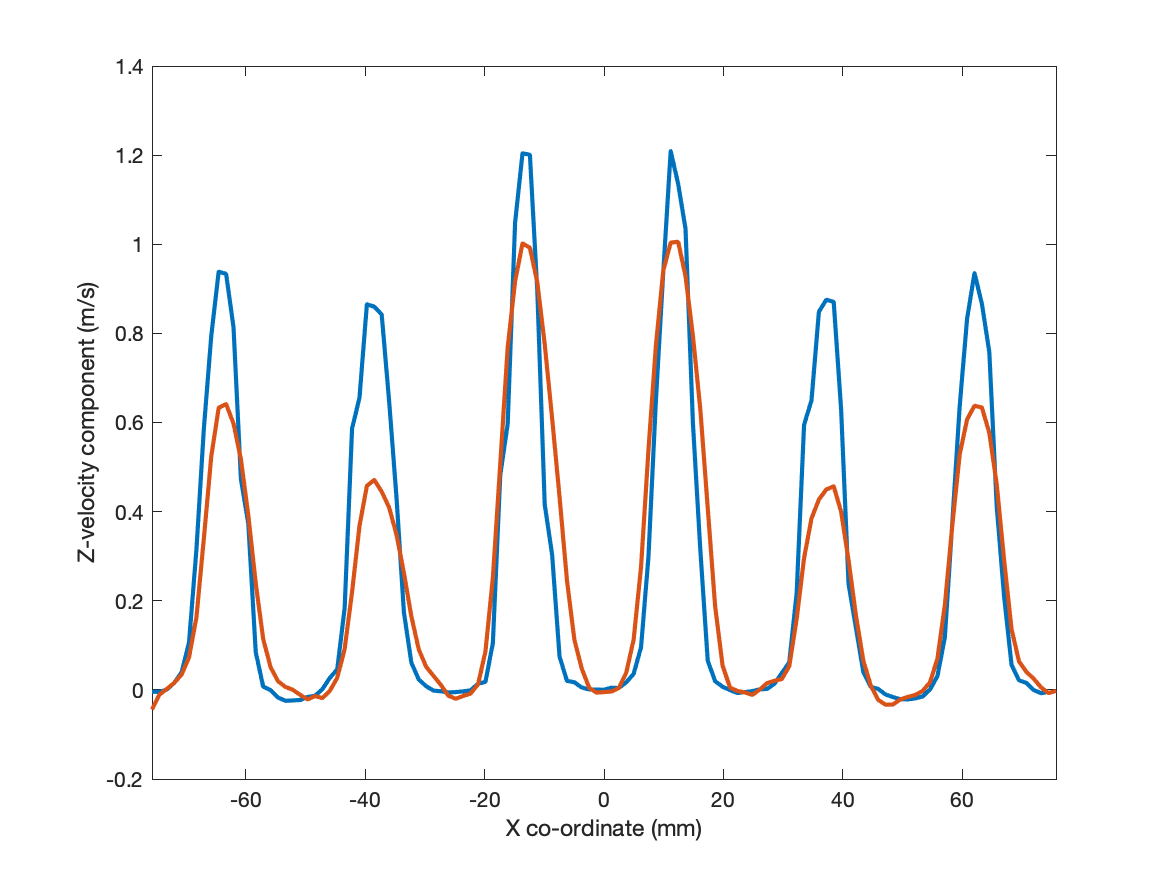}}\\
    \subcaptionbox{Red line $\mathrm{Re}_\mathrm{p}$ = 400\label{fig:Lineplot_red_Rep400SUC}}{\includegraphics[width=6cm]{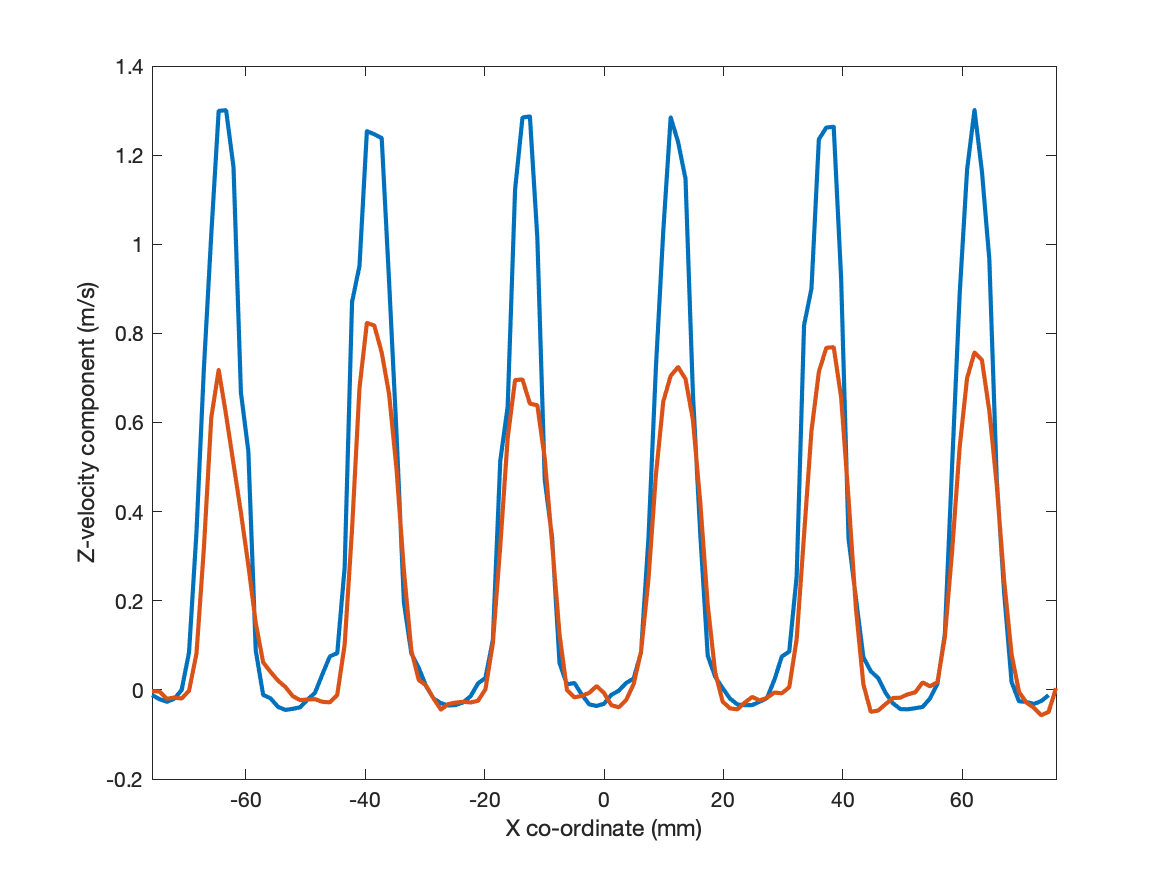}}
	\subcaptionbox{Black line $\mathrm{Re}_\mathrm{p}$ = 400\label{fig:Lineplot_black_Rep400SUC}}{\includegraphics[width=6cm,clip = 0cm 0cm 0cm 0.0cm]{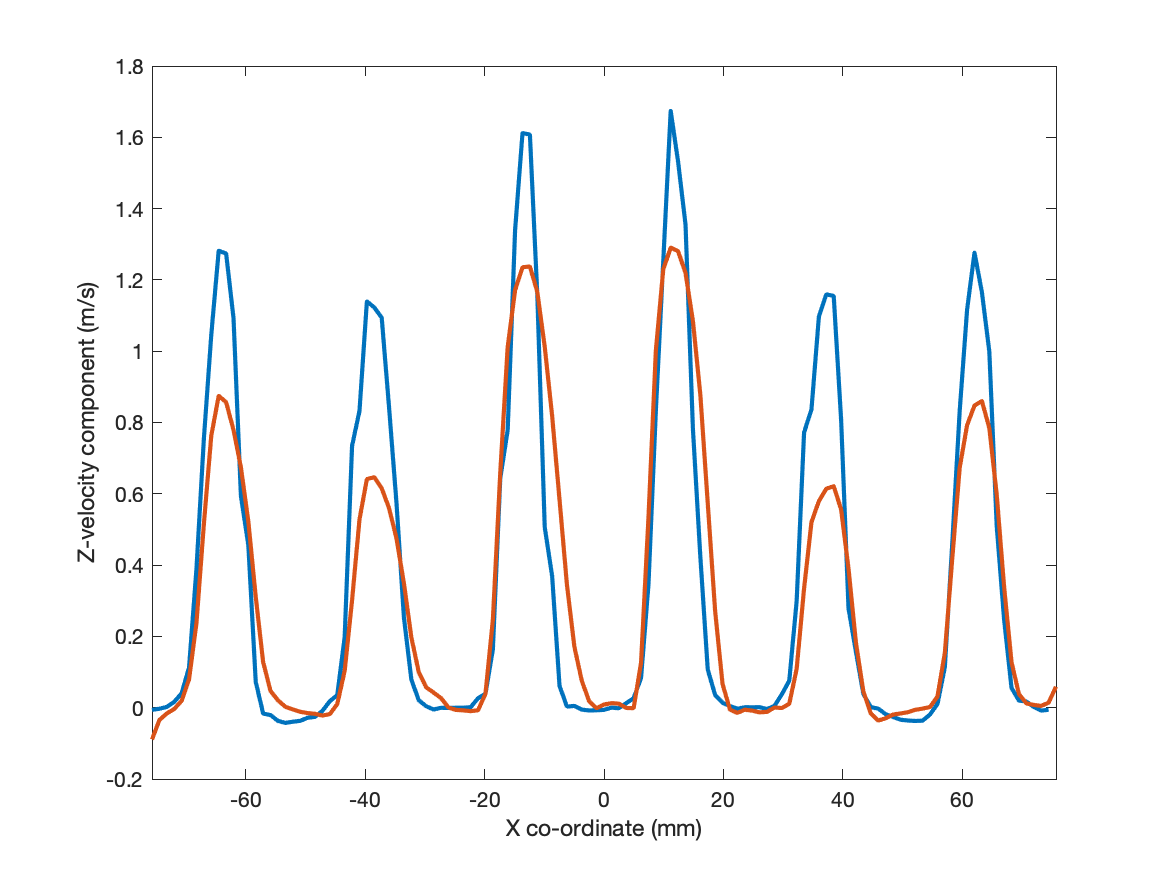}}\\
	\caption{Comparison of the numerical and experimental averaged axial velocity ($\overline{V_z}$) for the SUC packing along the red line (periphery of packed bed) and black line (centre of packed bed) marked in (a,b) for (c,d) $\mathrm{Re}_\mathrm{p}$ = 200, (e,f) $\mathrm{Re}_\mathrm{p}$ = 300, and (g,h) $\mathrm{Re}_\mathrm{p}$ = 400.}
	\label{fig:LineplotsSUC}
\end{figure}

First, the observations from contour plots from the experimental data, shown in Figures \ref{fig:AvgVzSUC200Exp}, \ref{fig:AvgVzSUC300Exp}, and \ref{fig:AvgVzSUC400Exp} are discussed. The velocity of the jets is maximum for the four central pores near to which the central jet is placed at the bottom of the setup. These jets from four central pores are highlighted with a dashed rectangle in the contour plot from the experimental data for $\mathrm{Re}_\mathrm{p}$ = 200 (see Figure \ref{fig:AvgVzSUC200Exp}). This indicates that the influence of the presence of the central jet is preserved, even after the flow passed through 18 layers of the SUC particle bed. It is observed, that, even though the flow has laterally dispersed to the pores at the periphery, the axial velocity through pores differs considerably from the central to the peripheral region of the packed particle bed. Moreover, as expected, the velocity of each individual pore jet is the highest at its centre, and gradually reduces towards its periphery \citep{pope2000turbulent}. The pore jets seem to be elliptical near the central region of the packed bed, while they are completely transformed into rectangular shape near the periphery of the packed bed. However, before transforming into rectangular pore jets, there is a region where they appear as distorted ellipses with their maximum velocities reduced by around 50\% compared to the pore jets from the four central pores. The group of elliptical pore jets is highlighted with dashed rectangles for $\mathrm{Re}_\mathrm{p}$ = 300 and 400 (see Figure \ref{fig:AvgVzSUC300Exp} and \ref{fig:AvgVzSUC400Exp}). This highlights that, even though the physical size and structure of each pore is identical, the local inlet flow conditions at the pores eventually dictate the velocity magnitude of each jet. Also, it can be noted qualitatively that the jet structure through each pore remains overall similar for all $\mathrm{Re}_\mathrm{p}$ conditions.

The flow structure is observed to be overall symmetric. Though the symmetrical nature of the flow structure is almost perfect about $X$ axis, there seems to be slight asymmetric profiles about $Y$ axis. For instance, comparing the jet structure of the four central pores, it can be observed that the central pores towards the right have a relatively large spatial extent of highest velocities, as compared to the central pores towards the left of the particle bed. Although, it is ensured that the central pipe and the entire setup are aligned as straight as possible in the vertical direction, the flow structures that cross through the layers of the packed bed are very sensitive to precise the setup alignment and any minute offset between the placement of packed bed units.

In the numerical results, the structure of the pore jets is not elliptical, but seems to preserve the shape of the pore from which they emerge. This is most clearly seen for the pore jets in the central region of the packed bed (see Figures \ref{fig:AvgVzSUC200Sim}, \ref{fig:AvgVzSUC300Sim}, and \ref{fig:AvgVzSUC400Sim}). The pore jets near the periphery of the packed bed tend to have a rectangular shape in both the experiments and the simulations. The overall pore jet structures from all the pores remain similar for different $\mathrm{Re}_\mathrm{p}$, which is predicted by the numerics as well as the experiments. The possible reason behind the discrepancy in the pore jet structure between the simulation and the experiment might be explained as follows: The region of interest in the present study is 152.5 $\times$ 152.5 mm, which is relatively large, and the spatial resolution is rather low (i.e. interrogation window size is 1.25 mm) compared to typical PIV studies for packed beds \citep{khayamyan2017transitionalSPIV,patil2013flow, Neeraj2023}. This is a critical issue in the experiments, especially at the jets boundaries, where velocity gradients are large so that variations in the velocity field over a small length scale are prone to average the velocity field. Hence, the sharpness of fluid flow structures observed in the contour plots obtained from the experiments tends to diffuse, whereas in the numerical predictions this is not the case, at least not to the same extent.

Figure \ref{fig:LineplotsSUC} shows the comparison between the numerical and experimental averaged velocity profile along the centre and periphery of the packed bed for all considered $\mathrm{Re}_\mathrm{p}$, indicated by black and red lines in Figures \ref{fig:ContourLinesSimSUC} and \ref{fig:ContourLinesExpSUC}, respectively. It can be observed that the maximum velocities at the periphery of the packed bed (see Figures \ref{fig:Lineplot_red_Rep200SUC}, \ref{fig:Lineplot_red_Rep300SUC}, and \ref{fig:Lineplot_red_Rep400SUC}), marked by red lines in Figures \ref{fig:ContourLinesSimSUC} and \ref{fig:ContourLinesExpSUC}, remain almost constant for all the pores at all $\mathrm{Re}_\mathrm{p}$. This is similar for both experimental and numerical studies. However, the maximum velocities along the centreline of the packing (see Figures \ref{fig:Lineplot_black_Rep200SUC}, \ref{fig:Lineplot_black_Rep300SUC} and \ref{fig:Lineplot_black_Rep400SUC}), as indicated by the black lines in Figures \ref{fig:ContourLinesSimSUC} and \ref{fig:ContourLinesExpSUC} vary considerably for all pores. Some interesting observations about the distribution of the average velocity can be made from the experimental results. For the experiments at all considered $\mathrm{Re}_\mathrm{p}$, the maximum velocity decreases and then increases again as one moves from central pores to peripheral pores, see Figures \ref{fig:Lineplot_black_Rep200SUC}, \ref{fig:Lineplot_black_Rep300SUC}, and \ref{fig:Lineplot_black_Rep400SUC}. A gradual reduction in the maximum velocity from the central to the peripheral pores is expected, as the velocities of the central jet decrease near the sides \citep{pope2000turbulent}, which  eventually  disperse towards the peripheral pores. The reason behind the increase of this maximum velocity at the periphery might be the complex interaction between the fluid flow and the different layers of the packed particle bed within the interstitial spaces of different spheres. This highlights that the flow near the peripheral pores experiences a contraction in the actual area available for the fluid flow, and hence the fluid velocity tends to increase. The numerical results (the black lines) also show a similar behaviour, although not as pronounced as observed in experiments. 

For all considered $\mathrm{Re}_\mathrm{p}$, the common observation is that the average fluid velocity is usually slightly lower in the experiments compared to predictions from the simulations. For the case with $\mathrm{Re}_\mathrm{p}$ = 200, the fluid velocity compares very well between the experiment and the simulation for the central pores, see Figure \ref{fig:Lineplot_black_Rep200SUC}. However, for the cases with $\mathrm{Re}_\mathrm{p}$ = 300 and 400, the deviation in the average fluid velocity magnitude between the experiments and the simulations increases together with the turbulence level in the flow. The deviation between the experimental and numerical results are attributed to the following reasons: The surfaces of the spheres in the experiments are relatively rough due to their production method, whereas the surfaces of the spheres in the simulations are completely smooth. This difference leads to a different behaviour of the boundary layers on the particle surfaces, which can lead to a reduction in velocity magnitude in case of experiments. Another reason why the velocity magnitudes are consistently higher in the simulations compared to the experimental results is summarized in Table \ref{table:VolumeComp}. In this table, the volumetric flow rate obtained by SPIV is 9 - 14 \% lower than the flow rate based on the particle Reynolds number. This difference in volumetric flow rate does not exist in the numerical simulation. 

\begin{figure}[h!]
	\centering
	\subcaptionbox{$\mathrm{Re}_\mathrm{p}$ = 200\label{fig:AvgVzHistSUC200SimExpt}}{\includegraphics[width=8cm]{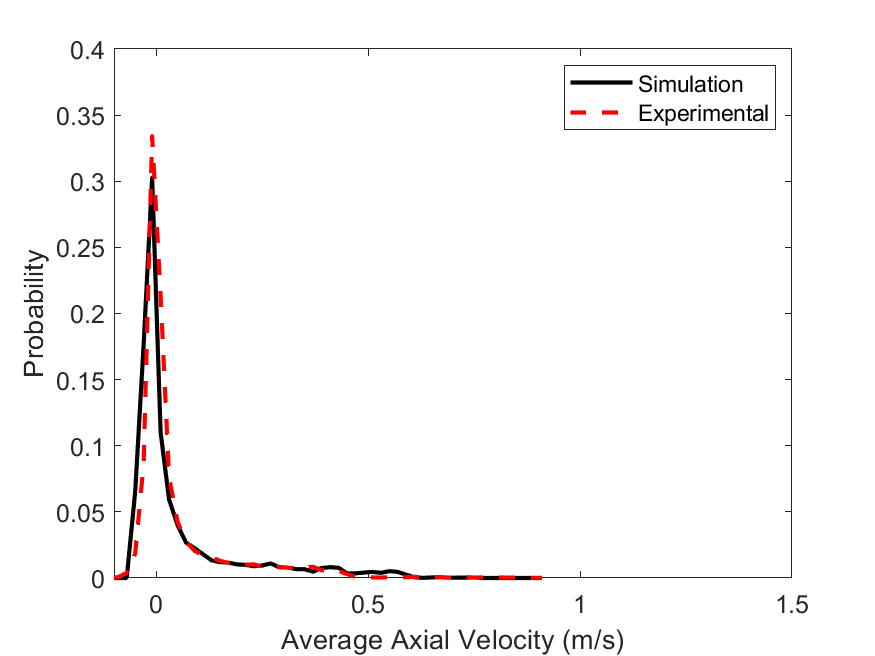}}
    \subcaptionbox{$\mathrm{Re}_\mathrm{p}$ = 300\label{fig:AvgVzHistSUC300SimExpt}}{\includegraphics[width=8cm]{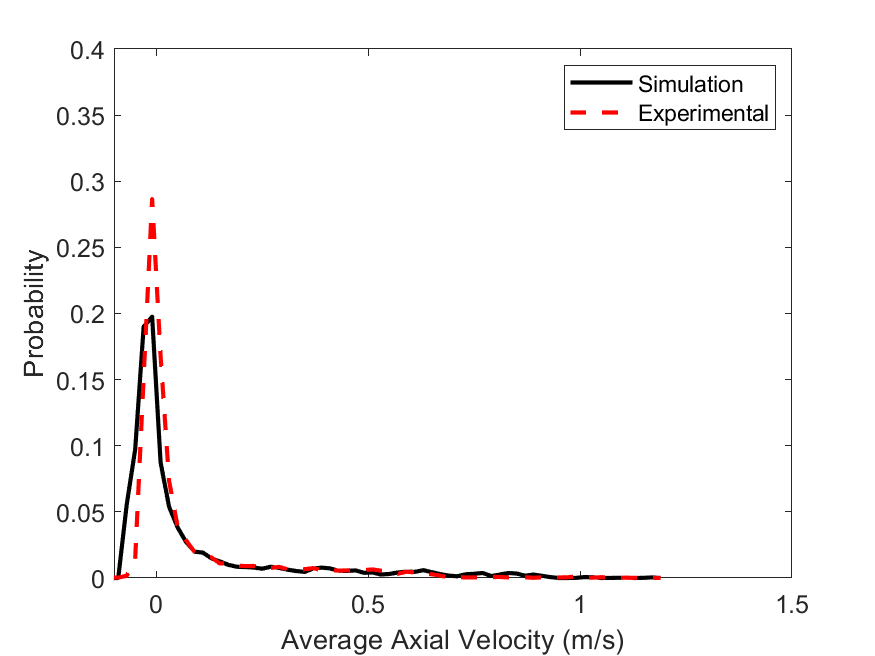}}\\	
    \subcaptionbox{$\mathrm{Re}_\mathrm{p}$ = 400\label{fig:AvgVzHistSUC400Sim}}{\includegraphics[width=8cm]{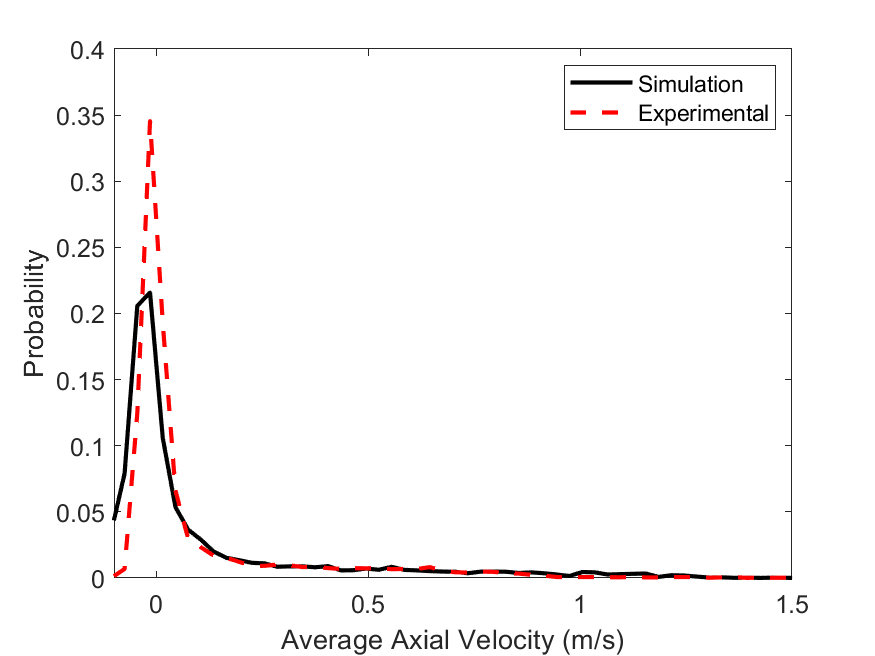}}\\
	\caption{Comparison of numerical and experimental distribution plots for averaged axial velocity ($\overline{V_z}$) for the SUC packing at (a) $\mathrm{Re}_\mathrm{p}$ = 200, (b) $\mathrm{Re}_\mathrm{p}$ = 300, and (c) $\mathrm{Re}_\mathrm{p}$ = 400.}
	\label{fig:AvgVzHistSUCComp}
\end{figure}

Figure \ref{fig:AvgVzHistSUCComp} shows the fluid velocity probability distributions for the velocity profiles shown in Figure \ref{fig:AvgVzSUCComp}. The probability is obtained by dividing the corresponding number of fluid velocity vectors by the total number of velocity vectors. These velocity distribution plots provide quantitative information about the distribution of the mean velocities in the measurement plane. It can be observed that there is good agreement between the numerical and experimental results for all considered $\mathrm{Re}_\mathrm{p}$, especially for $\mathrm{Re}_\mathrm{p}$ = 200. It is interesting to note, that even though flow structures and velocity magnitudes at some spatial locations differ between the experimental and simulation results (see Figures \ref{fig:AvgVzSUCComp} and \ref{fig:LineplotsSUC}), a good agreement is achieved as far as the distribution of the average axial velocity at the entire plane of measurement is considered. For all considered $\mathrm{Re}_\mathrm{p}$, the distributions peak at 0 m/s, and indicate that nearly 20-35\% of the velocity vectors in the measurement plane have zero velocity. The probability is around 5\% or less for most of the non-zero velocities values in the flow field. The maximum average velocities increase from around 0.9 m/s to 1.5 m/s for the experiments and simulations as $\mathrm{Re}_\mathrm{p}$ increases. It can be observed that the width of the probability distribution plots remain similar as $\mathrm{Re}_\mathrm{p}$ increases. For instance, this can be observed for a probability of around 2.5 \% where the velocities are in the narrow range of -0.05 to 0.1 m/s for all $\mathrm{Re}_\mathrm{p}$. Therefore, the higher range of velocities are significantly affected when the $\mathrm{Re}_\mathrm{p}$ is increased.

\subsection{Flow characteristics for the BCC packed bed}\label{subsec:RD_BCC}

\begin{figure}[h!]
	\centering
	\subcaptionbox{Simulation $\mathrm{Re}_\mathrm{p}$ = 200\label{fig:AvgVzBCC200Sim}}{\includegraphics[width=8cm]{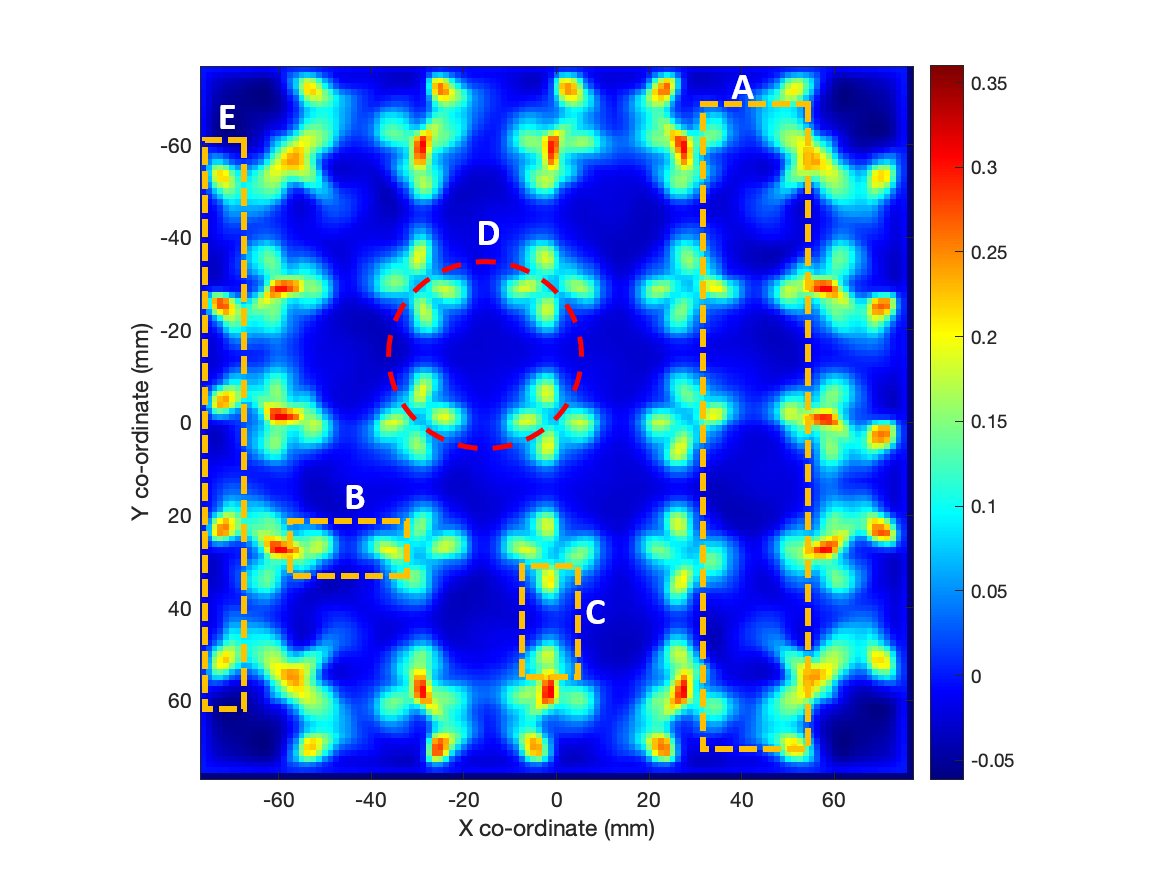}}
	\subcaptionbox{Experiment $\mathrm{Re}_\mathrm{p}$ = 200\label{fig:AvgVzBCC200Exp}}{\includegraphics[width=8cm,clip = 0cm 0cm 0cm 0.0cm]{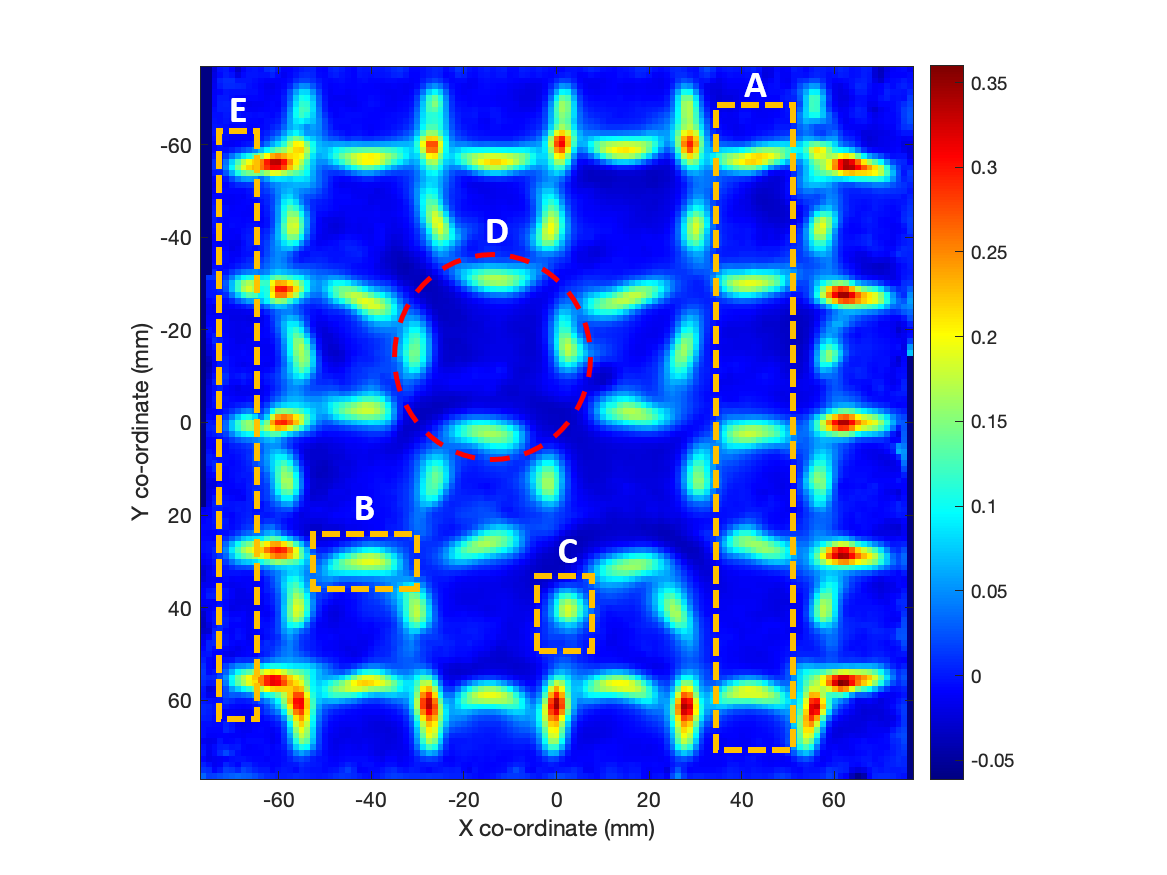}}\\
    \subcaptionbox{Simulation $\mathrm{Re}_\mathrm{p}$ = 300\label{fig:AvgVzBCC300Sim}}{\includegraphics[width=8cm]{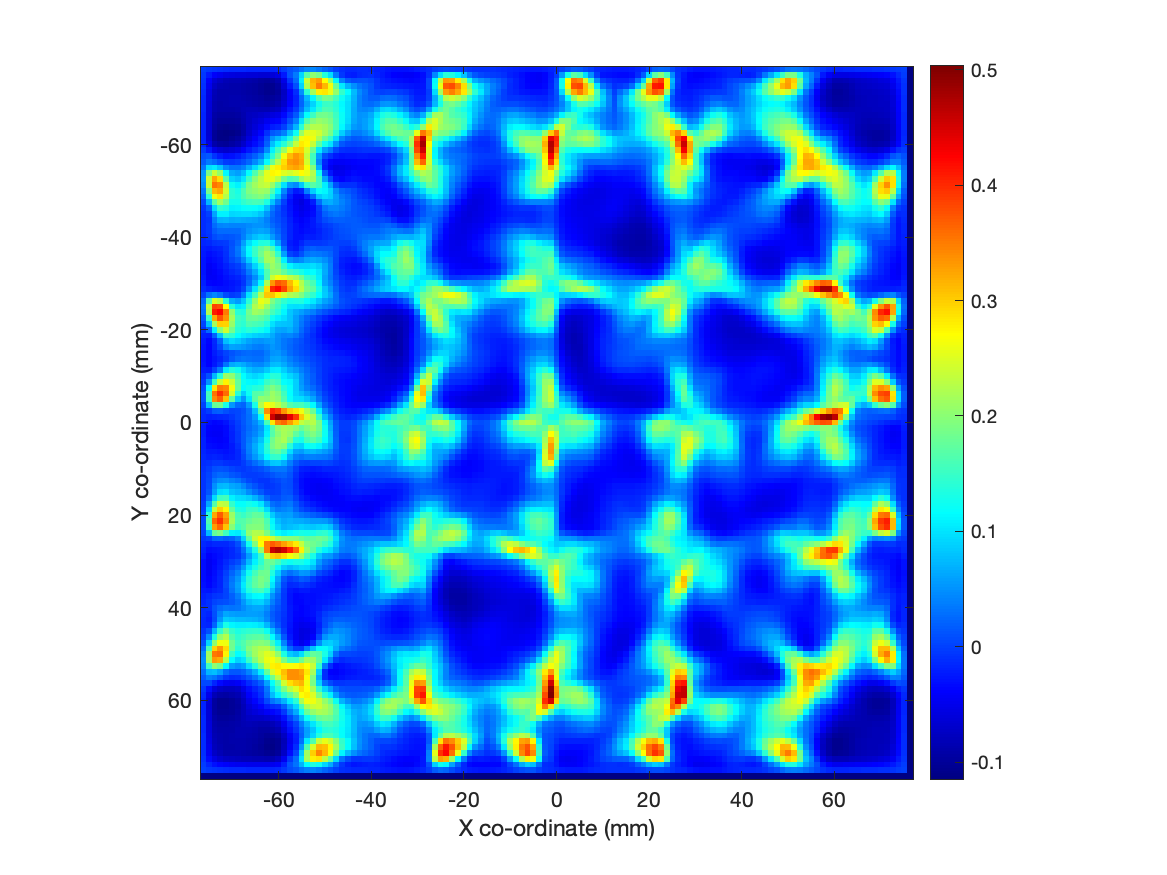}}
	\subcaptionbox{Experiment $\mathrm{Re}_\mathrm{p}$ = 300\label{fig:AvgVzBCC300Exp}}{\includegraphics[width=8cm,clip = 0cm 0cm 0cm 0.0cm]{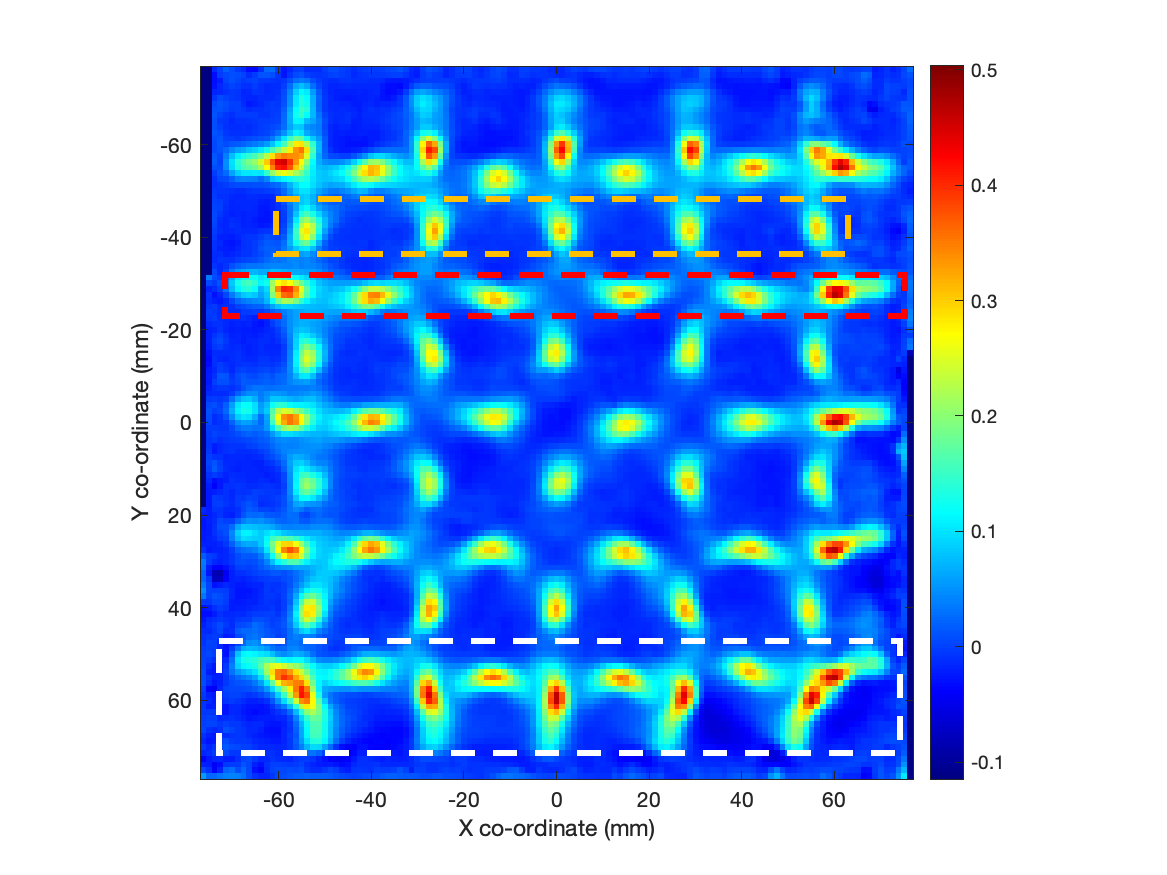}}\\
    \subcaptionbox{Simulation $\mathrm{Re}_\mathrm{p}$ = 400\label{fig:AvgVzBCC400Sim}}{\includegraphics[width=8cm]{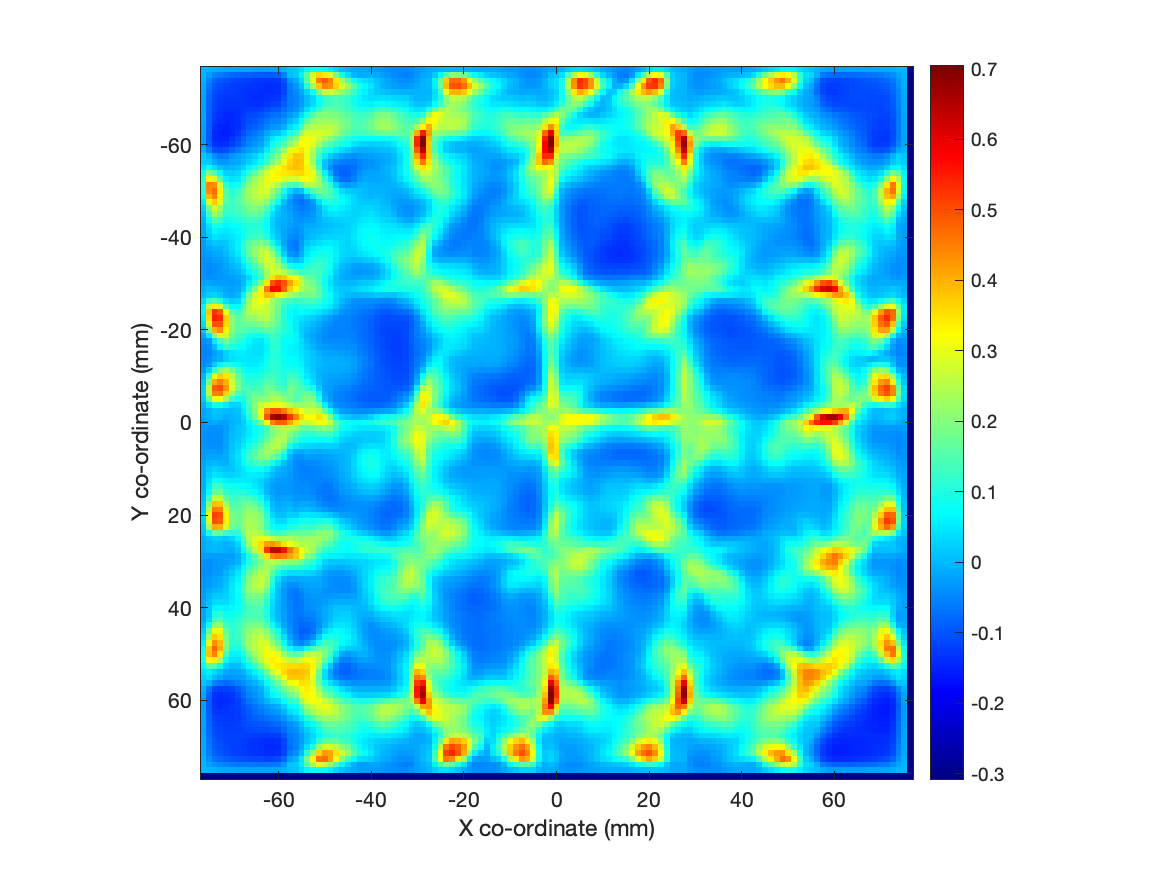}}
	\subcaptionbox{Experiment $\mathrm{Re}_\mathrm{p}$ = 400\label{fig:AvgVzBCC400Exp}}{\includegraphics[width=8cm,clip = 0cm 0cm 0cm 0.0cm]{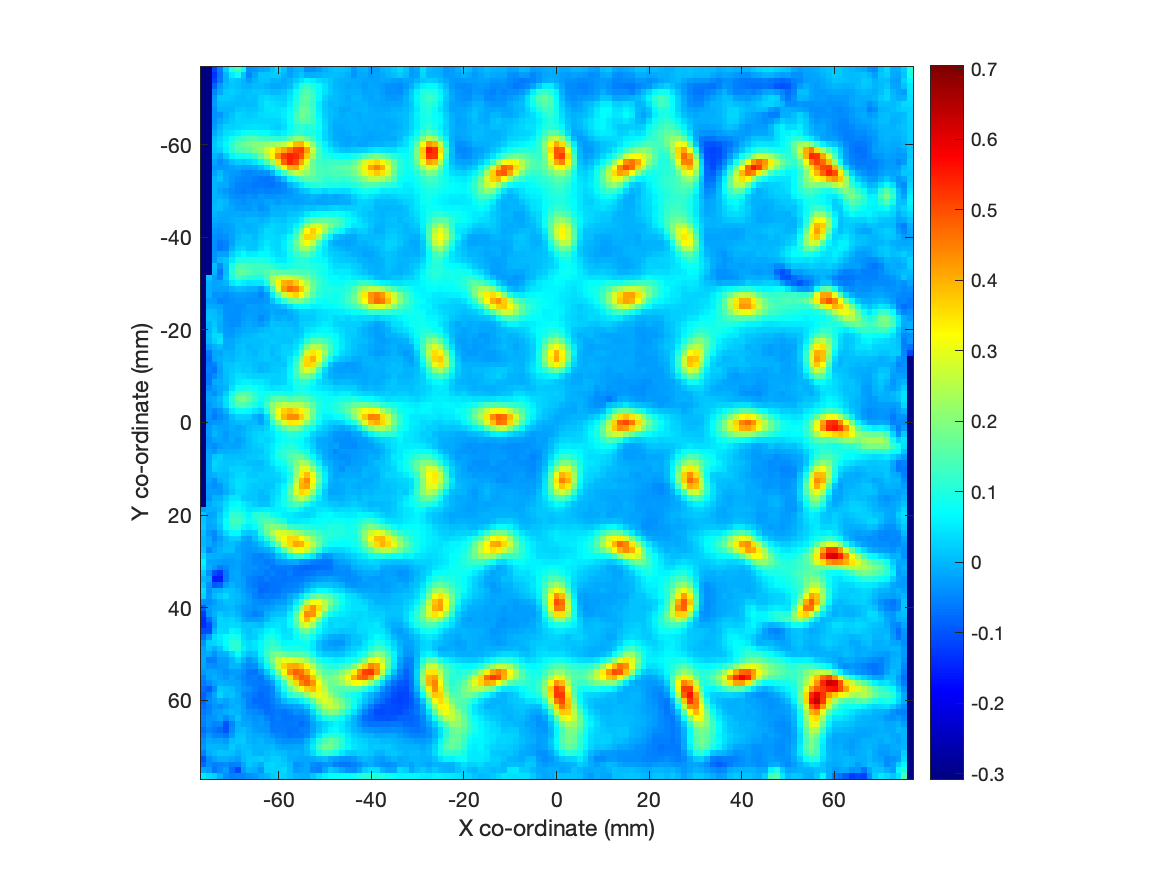}}\\
	\caption{Comparison of numerical and experimental contour plots of averaged axial velocity ($\overline{V_z}$) for BCC packing at (a,b) $\mathrm{Re}_\mathrm{p}$ = 200, (c,d) $\mathrm{Re}_\mathrm{p}$ = 300, and (e,f) $\mathrm{Re}_\mathrm{p}$ = 400.}
	\label{fig:AvgVzBCCComp}
\end{figure}

Figure \ref{fig:AvgVzBCCComp} compares the contour plots of the average axial fluid velocity between the experiments and simulations for the BCC packed beds for $\mathrm{Re}_\mathrm{p}$ = 200, 300, and 400. First, the observations from contour plots from the experimental data, shown in Figures \ref{fig:AvgVzBCC200Exp}, \ref{fig:AvgVzBCC300Exp}, and \ref{fig:AvgVzBCC400Exp} are discussed. It is noted that there are 60 pores in total for the BCC packed bed, and that the pore size is smaller compared to the SUC packed bed, see Figure \ref{fig:SUC-BCC-structure}. Out of the 60 pore jets, 56 jets can be clearly observed in the contour plot, while the remaining pore jets, at the four corners of the packed bed, have significant interactions with the adjacent jets and thus overlapped with them. For the case with $\mathrm{Re}_\mathrm{p}$ = 400, the maximum velocity of the pore jets is observed to be comparable between the central and peripheral pores, but for $\mathrm{Re}_\mathrm{p}$ = 200 and 300, differences in maximum velocities are observed. For all $\mathrm{Re}_\mathrm{p}$, it is observed qualitatively that the spatial extent of higher pore jet velocities is larger for the peripheral pores as compared to central pores. This suggests that for the same $\mathrm{Re}_\mathrm{p}$, the fluid flow disperses significantly in the lateral directions in the BCC packing, which is less pronounced in the SUC packing. It is noted that there are complex interactions of the fluid with the spheres of the packed bed for the BCC packing as fluid penetrates all full and weak layers of the packing and also disperses in lateral $X$ and $Y$ directions.

\begin{figure}[h!]
	\centering
	\subcaptionbox{Simulation \label{fig:ContourLinesSimBCC}}{\includegraphics[width=5cm]{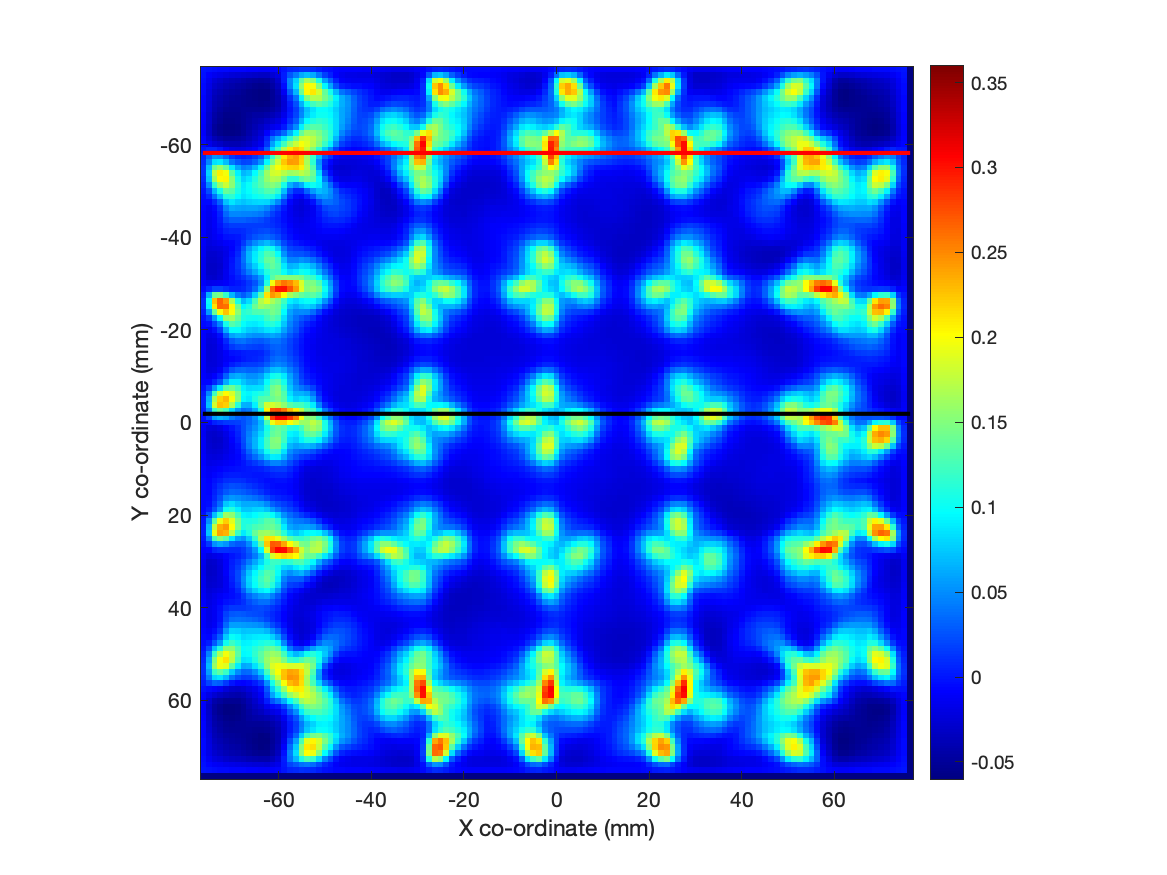}}
	\subcaptionbox{Experiment \label{fig:ContourLinesExpBCC}}{\includegraphics[width=5cm,clip = 0cm 0cm 0cm 0.0cm]{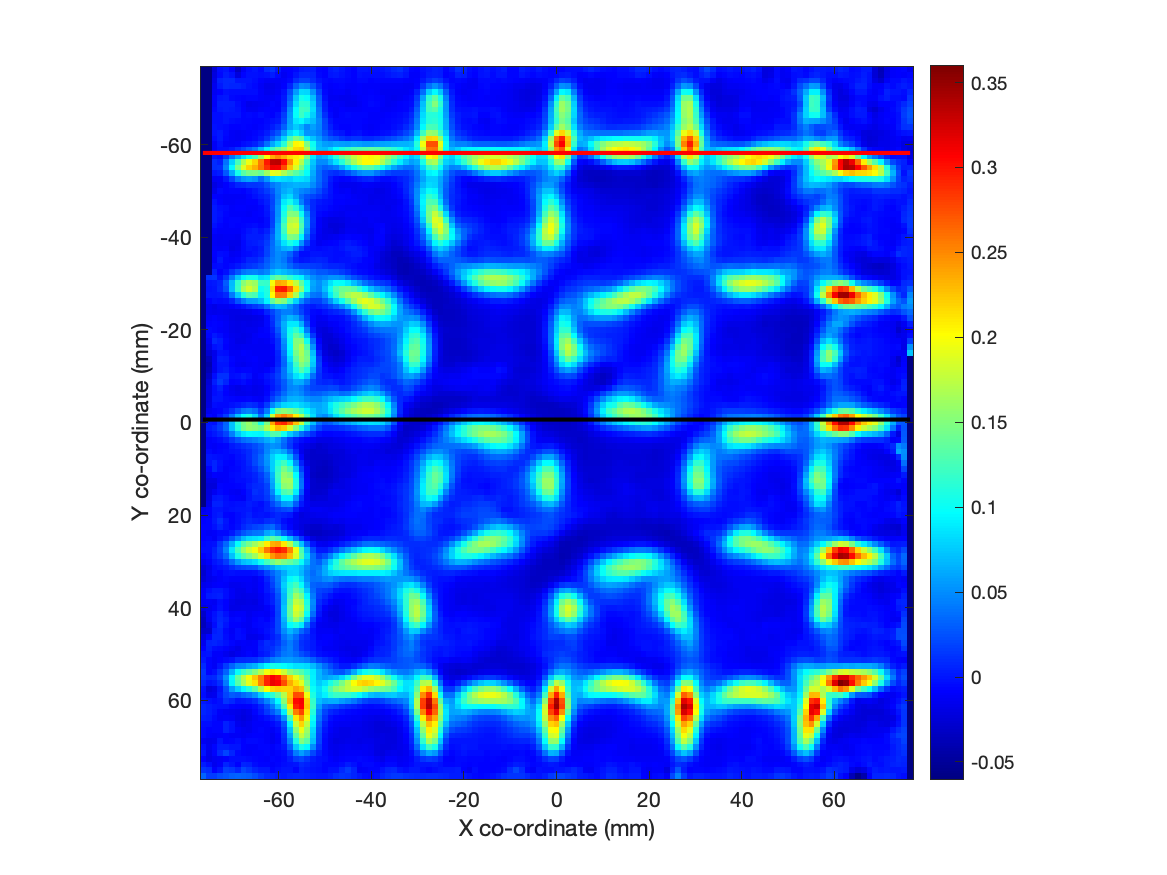}}\\
	\subcaptionbox{Red line $\mathrm{Re}_\mathrm{p}$ = 200\label{fig:Lineplot_red_Rep200BCC}}{\includegraphics[width=6cm]{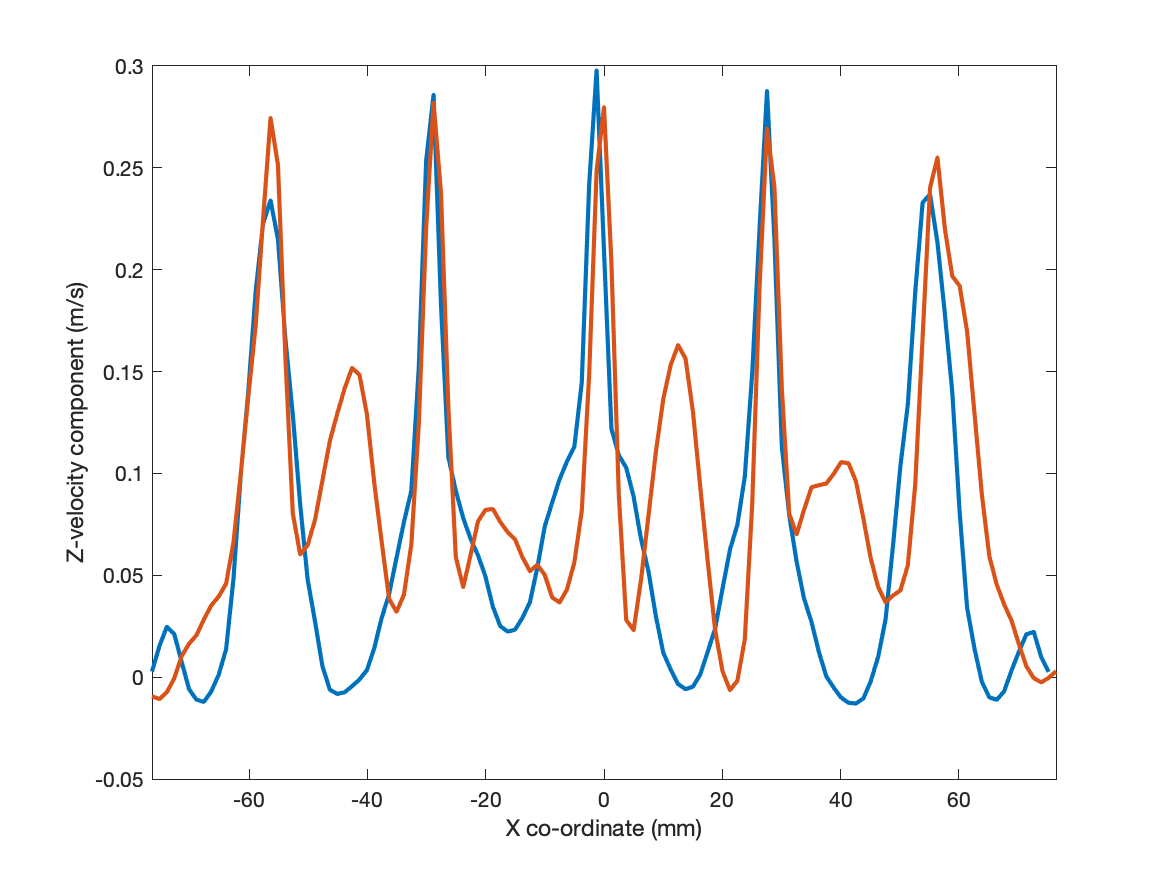}}
	\subcaptionbox{Black line $\mathrm{Re}_\mathrm{p}$ = 200\label{fig:Lineplot_black_Rep200BCC}}{\includegraphics[width=6cm,clip = 0cm 0cm 0cm 0.0cm]{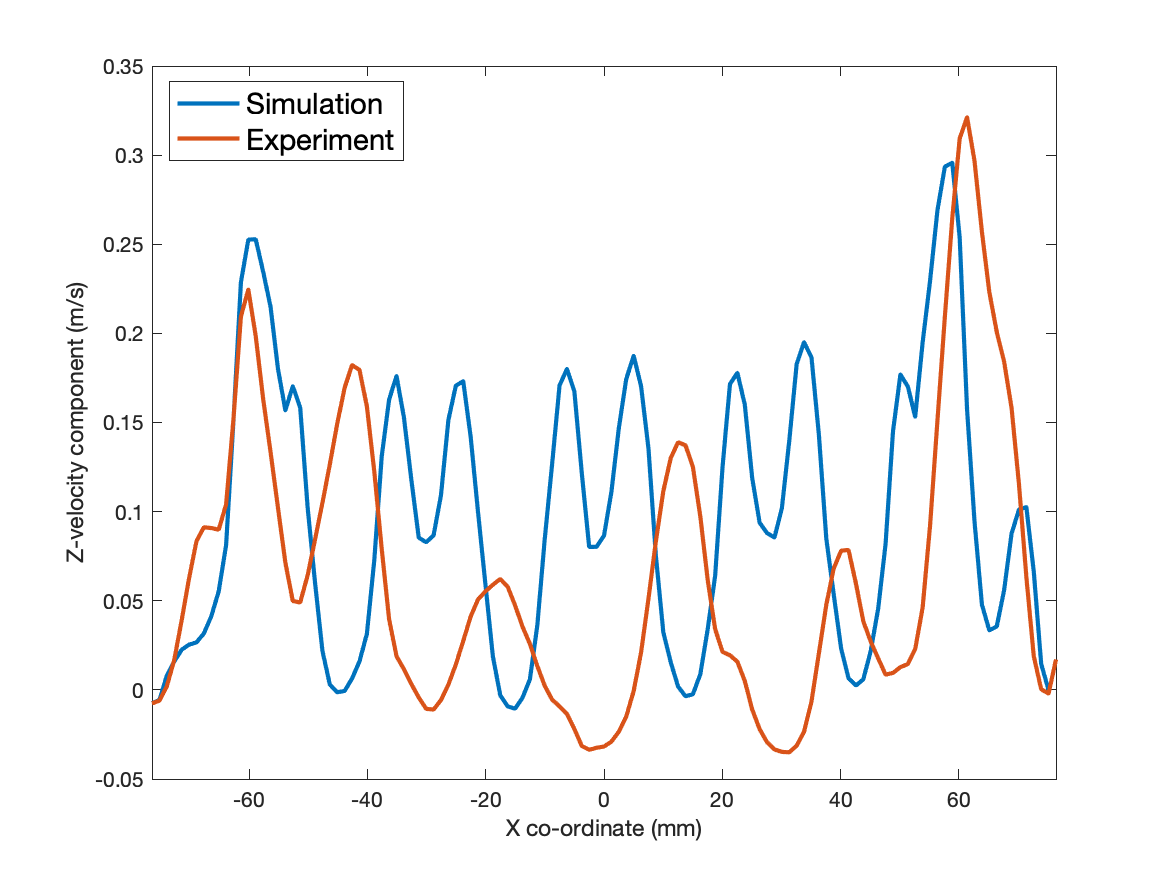}}\\
    \subcaptionbox{Red line $\mathrm{Re}_\mathrm{p}$ = 300\label{fig:Lineplot_red_Rep300BCC}}{\includegraphics[width=6cm]{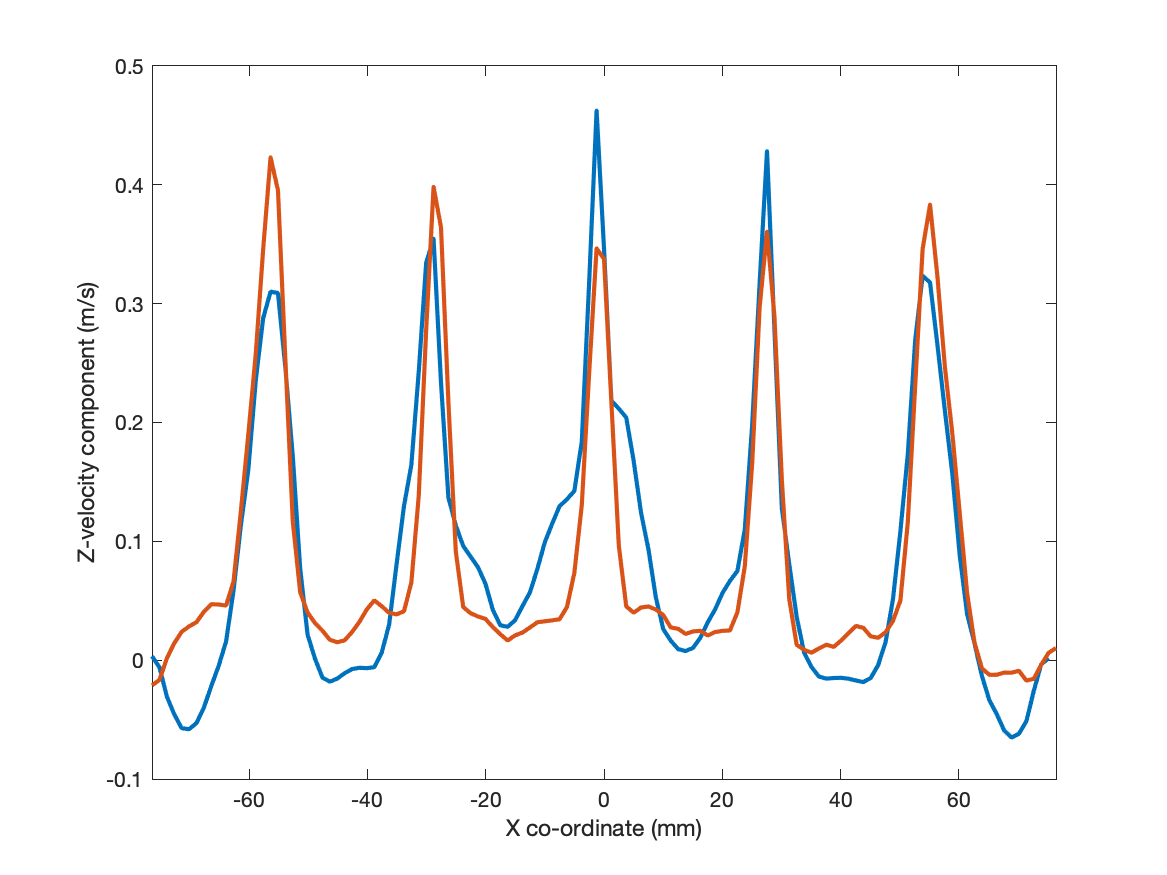}}
	\subcaptionbox{Black line $\mathrm{Re}_\mathrm{p}$ = 300\label{fig:Lineplot_black_Rep300BCC}}{\includegraphics[width=6cm,clip = 0cm 0cm 0cm 0.0cm]{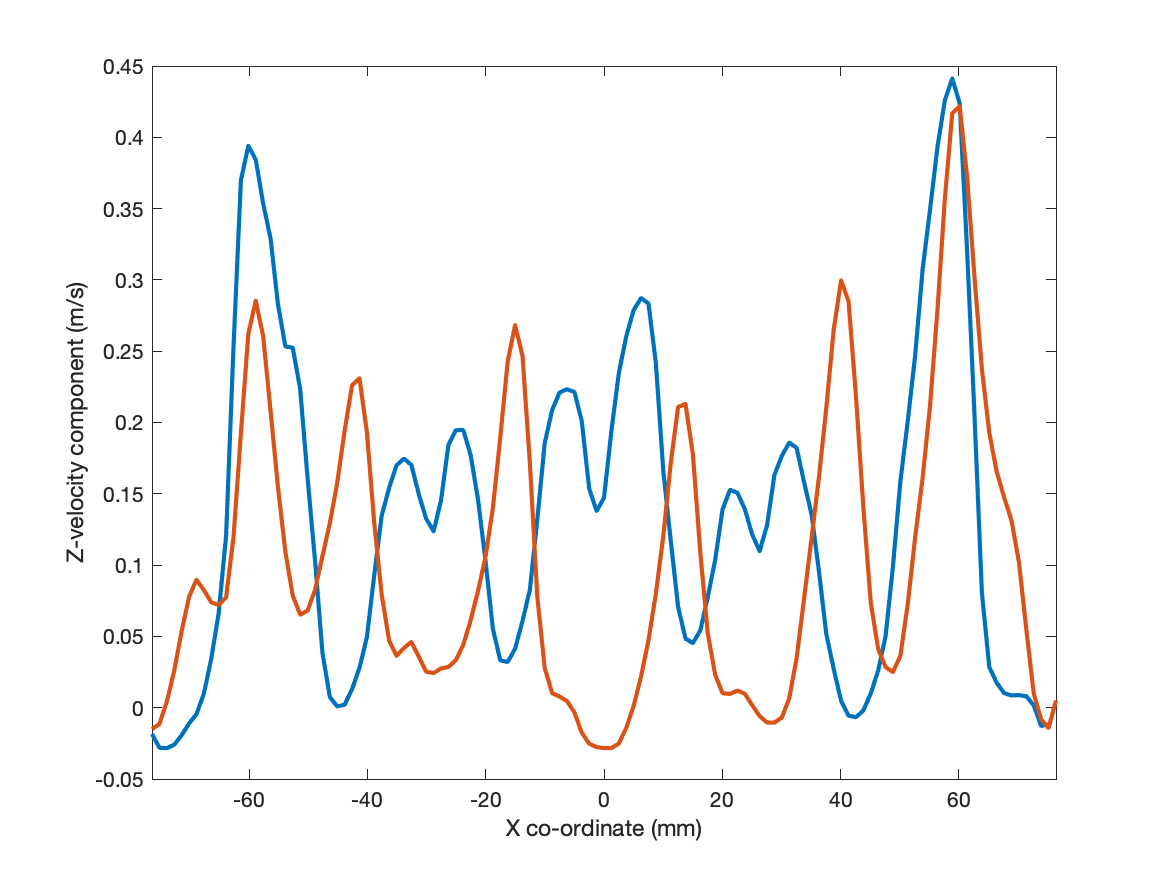}}\\
    \subcaptionbox{Red line $\mathrm{Re}_\mathrm{p}$ = 400\label{fig:Lineplot_red_Rep400BCC}}{\includegraphics[width=6cm]{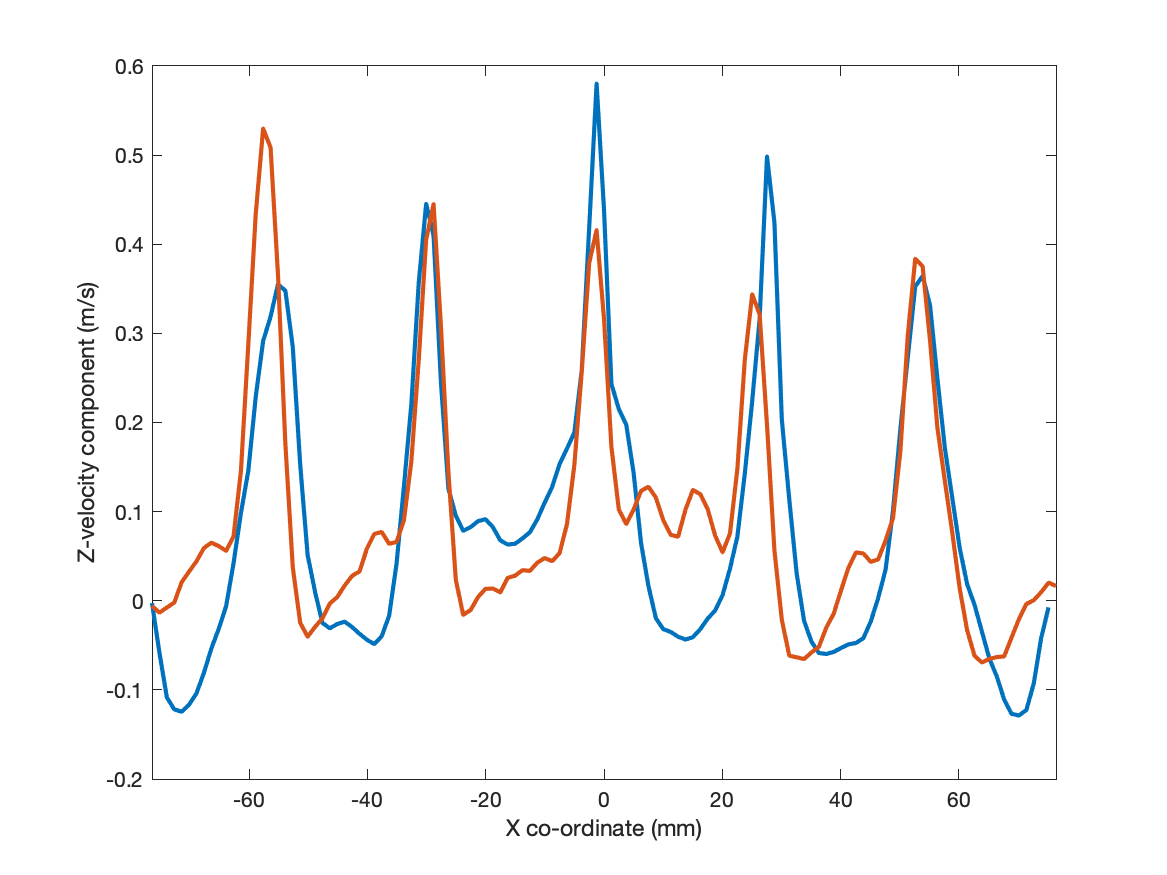}}
	\subcaptionbox{Black line $\mathrm{Re}_\mathrm{p}$ = 400\label{fig:Lineplot_black_Rep400BCC}}{\includegraphics[width=6cm,clip = 0cm 0cm 0cm 0.0cm]{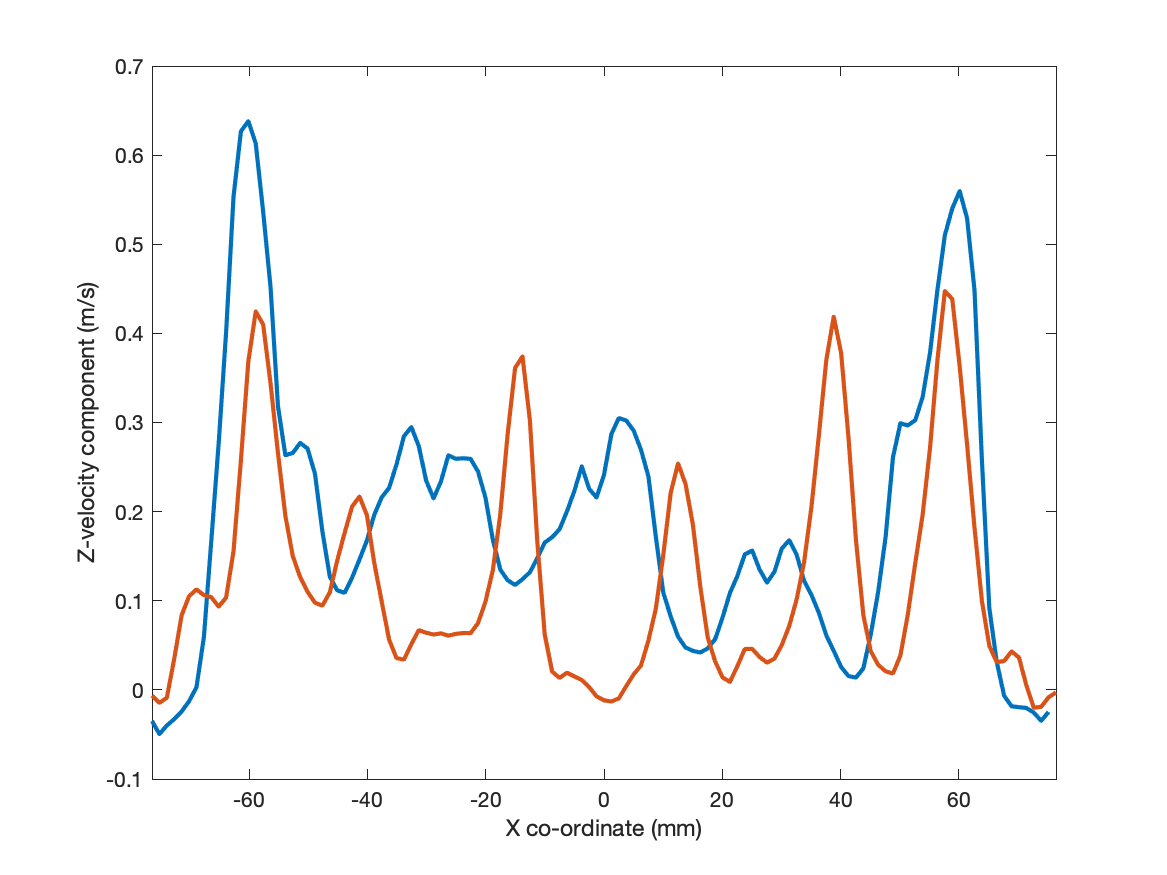}}\\
	\caption{Comparison of numerical and experimental averaged axial velocity ($\overline{V_z}$) for the BCC packing along the red line (periphery of packed bed) and the black line (centre of packed bed) marked in (a,b) for (c,d) $\mathrm{Re}_\mathrm{p}$ = 200, (e,f) $\mathrm{Re}_\mathrm{p}$ = 300, and (g,h) $\mathrm{Re}_\mathrm{p}$ = 400.}
	\label{fig:LineplotsBCC}
\end{figure}

To highlight the different features of the jets, dashed rectangles are drawn to focus the attention on the contour plot for $\mathrm{Re}_\mathrm{p}$ = 300 in Figure \ref{fig:AvgVzBCC300Exp}. It is noted that, along a particular row or column, elliptical pore jets are oriented in the same direction, except for the rows or columns at the periphery. For instance, for the first dotted rectangle from the top (orange), all the elliptical pore jets have their major axis approximately oriented in the vertical direction, i.e. along the $Y$ axis. The second dashed rectangle (red) shows the pore jets oriented with the horizontal major axis, i.e. along the $X$ axis. The jet structure shown by the first and second dashed rectangles is observed for every consecutive row or column. The third dashed rectangle (white) groups the jets at the periphery, and it has alternate jets whose major axis is oriented in both the directions. Similar trends of the elliptical pore jet structures and their orientation are observed for $\mathrm{Re}_\mathrm{p}$ = 200 and 400 as well. For the case with the SUC packing, the pore jet structures and especially their orientation does not vary considerably for the different pores. However, the BCC packing produces a complex arrangement of jets, even though the considered $\mathrm{Re}_\mathrm{p}$ is the same. For $\mathrm{Re}_\mathrm{p}$ = 200, all elliptical pore jets are not distinctly visible along a particular row or column, which may be due to the lower momentum of the pore jets compared to the cases with $\mathrm{Re}_\mathrm{p}$ = 300 and 400. Hence, the pore jets may be more susceptible to surrounding perturbations in the velocity field, as they eject out of the packing. 

In general, it is observed that the main patterns of the flow structures are similar between the simulations and the experiments. For instance, the area indicated with  the dashed rectangle `A' in Figures \ref{fig:AvgVzBCC200Sim} and \ref{fig:AvgVzBCC200Exp} for $\mathrm{Re}_\mathrm{p}$ = 200 shows very similar features. This comparison is also very good for the cases with $\mathrm{Re}_\mathrm{p}$ = 300 and 400, but $\mathrm{Re}_\mathrm{p}$ = 200 is chosen as an example here. In both the experiments and the simulations, there are five rows over which flow is distributed. However, a single elliptical pore jet appears in the experiments. While there are two distinct pore jets visible for each row in the numerical studies. This can be seen in the dashed rectangles `B' and `C' in the figures.

Similar features of the pore jets' structures in the simulations and experiments are also shown by the circle `D' in the figure. Due to the lower spatial resolution in the SPIV experiments compared to the simulations, multiple pore jets are overlapping, and appear to form a single pore jet in the experimental results. This was not observed in the case of the results obtained for the case with the SUC packing, as the distance between the centre of each pore was on the order of the diameter of the sphere of the packing. But for the cases with the BCC packing, this distance is reduced to 0.70 times the diameter of the sphere, see Figure \ref{fig:SUC-BCC-structure} for a qualitative comparison. Furthermore, the size of individual pore jets itself is larger in the cases with the SUC packing as compared to the pore jets from the BCC packing, compare Figure \ref{fig:AvgVzBCC200Exp} for BCC, and Figure \ref{fig:AvgVzSUC200Exp} for SUC. The effect of this can be observed, for instance, in the pore  jet flow in the simulation results near the wall, as shown by dashed rectangle `E'. Such pore jet flows near the wall tend to get averaged out in the corresponding experimental locations, due to the lower spatial resolution in the experiments compared to the simulations. \
Figure \ref{fig:LineplotsBCC} shows the comparison between the average axial velocity for the simulation and the experiments along the black line (i.e. the centre) and red line (i.e. the periphery) of the packed bed (see Figures \ref{fig:ContourLinesSimBCC} and \ref{fig:ContourLinesExpBCC}), for the case with the BCC structured particle packing. It is observed, that the average velocities along the red line at the periphery of the packed bed shows overall a good comparison between experiments and simulations for the cases with $\mathrm{Re}_\mathrm{p}$ = 300 and 400. In contrast, for $\mathrm{Re}_\mathrm{p}$ = 200, there is some deviation between the experiments and the simulations,  especially for some locations such as $X \approx$ -40 or 40 mm. The velocity profile along a black line at the centre of the packed bed shows a higher difference in the velocities from the experimental and the numerical results at the central pores. For the peripheral pores, the comparison between the experiments and the simulations is generally very good. It is observed, that it is difficult to obtain a perfect agreement between the experiments and the simulations at all locations for the BCC packing, as the pore jets' structures vary considerably between the experiments and the simulations, as seen earlier in Figure \ref{fig:AvgVzBCCComp}.

\begin{figure}[h!]
	\centering
	\subcaptionbox{$\mathrm{Re}_\mathrm{p}$ = 200\label{fig:StdVzHistBCC200SimExpt}}{\includegraphics[width=8cm]{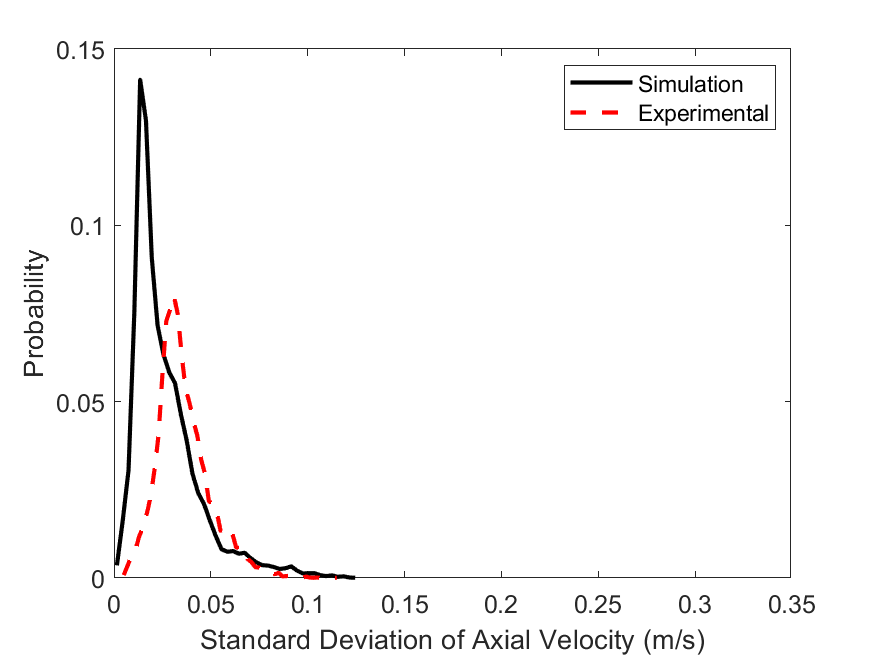}}
    \subcaptionbox{$\mathrm{Re}_\mathrm{p}$ = 300\label{fig:VzStdHistBCC300SimExpt}}{\includegraphics[width=8cm]{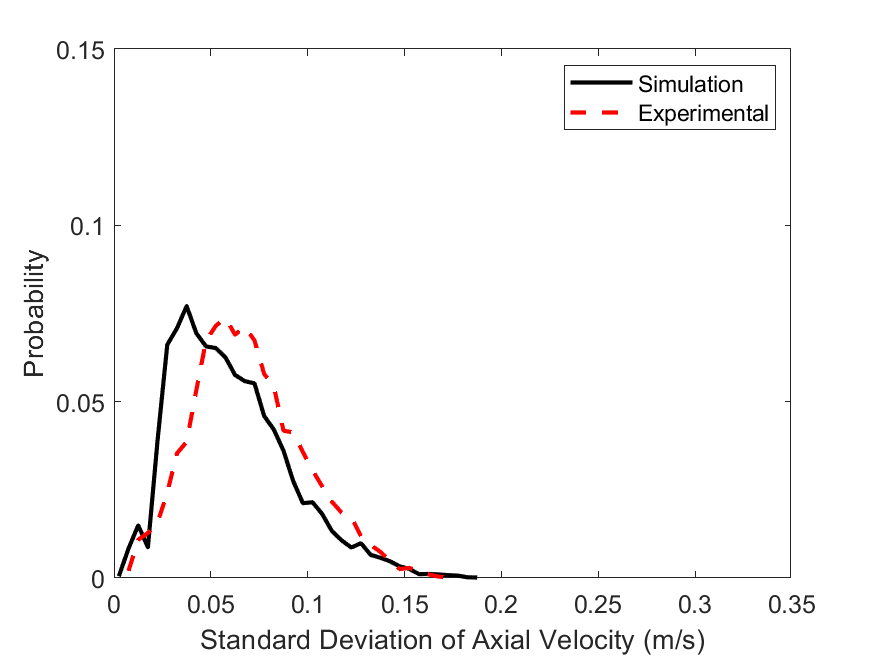}}\\	
	\caption{Comparison of numerical and experimental distribution plots for standard deviation of axial velocity for BCC packing at (a) $\mathrm{Re}_\mathrm{p}$ = 200, (b) $\mathrm{Re}_\mathrm{p}$ = 300.}
	\label{fig:StdVzHistBCCComp}
\end{figure}

A possible reason for some discrepancies observed between the experiments and the simulations for the BCC packing, can also be attributed to the surface finish of the particles in the particle packing. As mentioned above, the sphere surfaces are relatively rough in experiments, while they are completely smooth in simulations. The influence of surface roughness is amplified in the cases with the BCC packing compared to the SUC packing, as the number of spheres in the former is higher, and the distance between the spheres, the size of the interstitial pores, is lower in the BCC particle configuration. 
Hence, the fluid flow in the experiments experiences a slightly different effect in the boundary layers at the surfaces of the spheres compared to the simulations, and this difference is expected to be more pronounced in the case with the BCC packing than the case with the SUC packing.
Moreover, the dispersion of the fluid flow through the layers of the packed bed in case of the BCC configuration involves higher complexity, due to the more complex arrangement of the spheres and the subsequent flow structures in the packing. The surface roughness also likely triggers more velocity fluctuations in the fluid flow, which leads to the generation of vortices and, hence, better mixing within the fluid flow. Accordingly, as can be seen in Figure \ref{fig:StdVzHistBCCComp}, the standard deviation of the axial fluid velocities is found to be higher for the experiments compared to the simulations. Because of these reasons, the BCC configuration is likely to be more sensitive to the surface roughness of the spheres in the packing, and hence leads to the observed differences in flow structure and velocities at the exit of the packed bed between the experiments and simulations. Moreover, due to the relatively low spatial resolution of the presented SPIV results and the underestimation of the volumetric flow rate by SPIV (see Table \ref{table:VolumeComp}), these issues might also contribute to the observed differences. As shown in Figure \ref{fig:LineplotsSUC}, the boundary or peripheral pore jets in the simulations tend to have a slightly higher velocity magnitude compared to the experiments, especially for the case with $\mathrm{Re}_\mathrm{p}$ = 400. The variation in results could also appear due to the difference in operating time between simulation and experiments. The experimental images were captured over a period of 133 seconds, while simulations ran for a period of 2 seconds of physical time. 

\begin{figure}[h!]
	\centering
	\subcaptionbox{$\mathrm{Re}_\mathrm{p}$ = 200\label{fig:AvgVzHistBCC200SimExpt}}{\includegraphics[width=8cm]{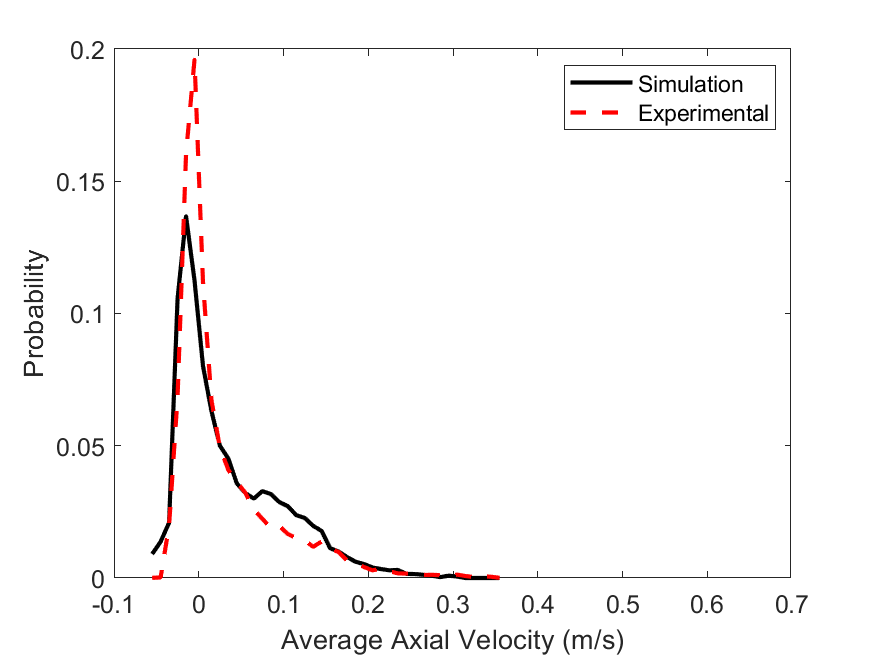}}
    \subcaptionbox{$\mathrm{Re}_\mathrm{p}$ = 300\label{fig:AvgVzHistBCC300SimExpt}}{\includegraphics[width=8cm]{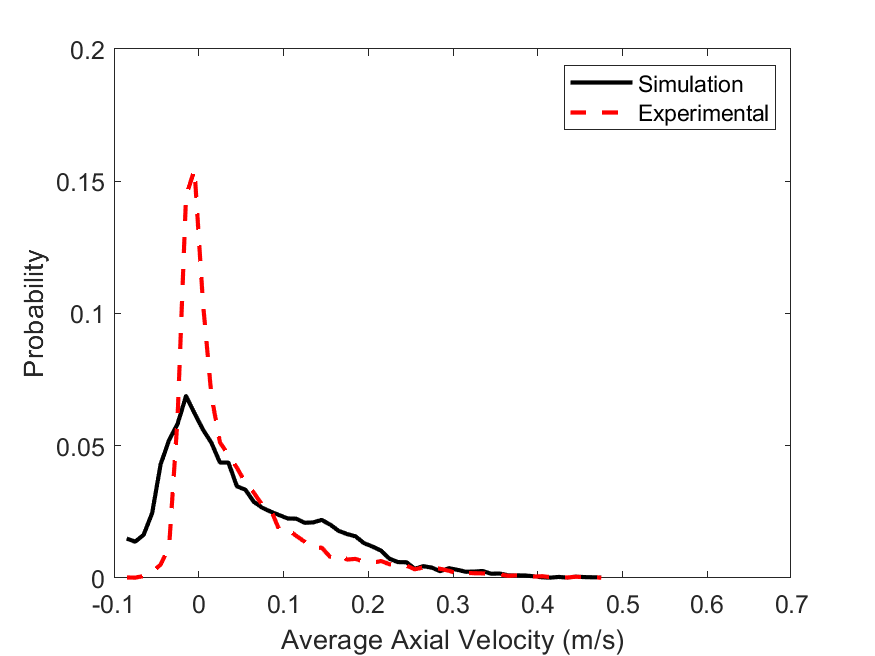}}\\	
    \subcaptionbox{$\mathrm{Re}_\mathrm{p}$ = 400\label{fig:AvgVzHistBCC400Sim}}{\includegraphics[width=8cm]{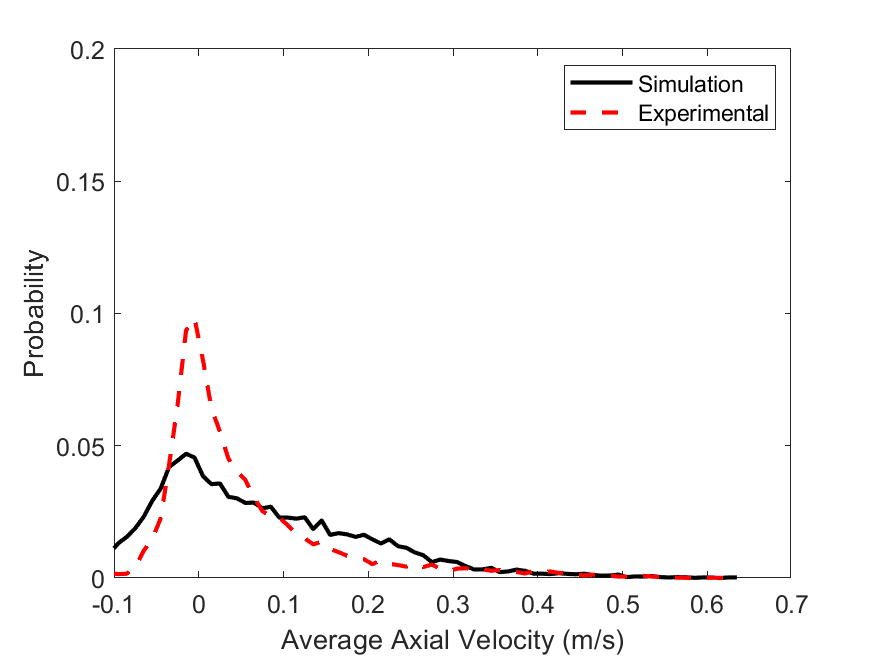}}\\
	\caption{Comparison of numerical and experimental distribution plots for averaged axial velocity ($\overline{V_z}$) for BCC packing at (a) $\mathrm{Re}_\mathrm{p}$ = 200, (b) $\mathrm{Re}_\mathrm{p}$ = 300, and (c) $\mathrm{Re}_\mathrm{p}$ = 400.}
	\label{fig:AvgVzHistBCCComp}
\end{figure}

Figure \ref{fig:AvgVzHistBCCComp} shows the probability distribution of the average velocities for the considered values of $\mathrm{Re}_\mathrm{p}$. Similar to the SUC configuration, the distribution shows a peak at 0 m/s, with probabilities in the range of 5-20 \%  for all $\mathrm{Re}_\mathrm{p}$. 
It can be observed, that the comparisons of the velocity distribution between the experiments and the simulations are better for $\mathrm{Re}_\mathrm{p}$ = 200 than for the two cases with $\mathrm{Re}_\mathrm{p}$ = 300 and 400. The positive axial velocities at which the probability reaches almost zero shows a good agreement between the simulations and experiments for all considered $\mathrm{Re}_\mathrm{p}$. For instance, the velocity increases from around 0.2 to 0.4 m/s as $\mathrm{Re}_\mathrm{p}$ is increased. Moreover, in contrast to the SUC configuration, the width of the probability distribution slightly increases for lower velocities with increasing $\mathrm{Re}_\mathrm{p}$. For instance, this is clearly observed by comparing the width of the distribution at a probability of 2.5 \%. The velocities vary in the range from -0.05 to 0.05 m/s for $\mathrm{Re}_\mathrm{p}$ = 200, and they vary from -0.05 to 0.1 m/s for $\mathrm{Re}_\mathrm{p}$ = 400. This shows that when increasing the value of $\mathrm{Re}_\mathrm{p}$ in case of the BCC packing, the lower and higher velocities of the flow field are increased. This is confirmed by the experiments as well as the simulations.

As can be seen from the contour plots in Figure \ref{fig:AvgVzBCCComp}, small negative velocities exist near the boundaries of the elliptically shaped pore jets. When $\mathrm{Re}_\mathrm{p}$ is increased, the magnitude of the negative 
velocities tends to increase and the distribution plot becomes slightly broader, as can be clearly observed in Figure \ref{fig:AvgVzHistBCCComp}. This implies that, as the pore jets' axial velocities increase with increasing $\mathrm{Re}_\mathrm{p}$, they induce a relatively strong re-circulation zone near the boundaries of the pore jets. The magnitude of the negative axial velocities is not observed to depend on $\mathrm{Re}_\mathrm{p}$ for the SUC configuration, as seen from Figure \ref{fig:AvgVzHistSUCComp}. For this case, the minimum negative velocity remains around -0.1 m/s for all $\mathrm{Re}_\mathrm{p}$. This is an interesting observation, confirmed by both the experiments and the simulations, even though the maximum axial velocity is always higher for the case of the SUC configuration as compared to the BCC configuration for the same $\mathrm{Re}_\mathrm{p}$.     

Figure \ref{fig:StdVzHistBCCComp} shows the comparison between the standard deviation of the axial fluid velocity between the experiments and the simulations for $\mathrm{Re}_\mathrm{p}$ = 200 and 300. As highlighted earlier, due to the surface roughness of the spheres, the velocity fields are more susceptible to fluctuations in the case of the experiments compared to the simulations. Hence, the peak in the distribution shifts towards higher values for the experiments for both $\mathrm{Re}_\mathrm{p}$, which is not seen to the same extent in the simulations.

\section{Conclusions}\label{sec:Conclusion}
This paper reports a detailed comparison between experimental and numerical studies for the fluid velocity field at the exit of simple unit cell (SUC) and body centered cubic (BCC) packed bed reactors, where a jet and a co-flow delivers the fluid flow from the bottom of the packed bed. Under the centre of the packing is a jet, and a co-flow is generated through a porous plate on which the particle packing rests. The objective is to study the details of the dispersion of the jet flow and co-flow through the layers of packed beds for flows with particle Reynolds numbers of $\mathrm{Re}_\mathrm{p}$ = 200, 300, and 400. Additionally, the paper provides model experimental results for the validation of simulation studies. The experimental setup involves 3D-printed spheres from which SUC and BCC packed beds are constructed. The experiments are performed with 18 layers of SUC and 25 layers of BCC (13 full layers and 12 weak layers). The effects of the wall flow are significantly reduced by having 3D-printed half spheres instead of full spheres at the channel wall and sphere interface. A co-flow of air is made to pass through a porous plate, so that a uniform co-flow velocity is obtained at the entrance of the packing, and the same boundary condition can be imposed in simulations. A fully developed jet flow is supplied at the exit of a central pipe, mounted directly under the centre of the packing. 

In the experiments, stereo particle image velocimetry (SPIV) is used in such a way, that the velocities over the entire region at the exit of the packed bed are obtained instantaneously.
The fluid flow of the jet is seeded with tracer particles, to enable the SPIV. 
The experiments are complex, as the out of plane component is the dominant velocity component of the flow field. Furthermore, the experimental measurements are carried out within the square channel, where the walls cause distortions in the recorded camera images. Therefore, the camera calibration is carefully executed and additionally a stereo self-calibration algorithm is used. The volumetric flow rate is observed to differ by 10-20 \% between the actual inlet flow rate, determined by mass flow controllers, and the flow rate achieved by integrating the results determined from the SPIV measurements. The present experimental SPIV arrangement helps to extract velocity field instantaneously over an entire region of the packed bed. 

For the numerical simulations, the immersed boundary method (IBM) with the direct forcing approach and an adaptively refined mesh is used. In the context of the flow simulations of fixed packed bed reactors, the proposed work shows that the IBM approach is highly suitable for modelling fixed packed bed reactors of any configuration. Especially the fact that the IBM can be straightforwardly applied to non-uniform and even moving packings of arbitrary shaped particles, with no significant increase in computational cost, is a great advantage compared to other simulation methodologies.

Overall, a good agreement between the simulations and the experimental results is observed.  Especially, for the SUC particle packing, there is generally a very good agreement between the experiments and the simulations for all considered particle Reynolds numbers.
However, the velocity magnitude was always slightly higher in the simulations than in the experiments. Interestingly, the axial velocity is found to increase at the pores in the packing at the exit, but this is more evident in experiments as compared to the simulations. 
The structure of the jets from the pores at the exit of the BCC configuration is found to differ between the experiments and simulation. In the simulations, for this configuration, mostly two jets appear from the pores, whereas only one jet appears from each pore in the experiments.
However, the overall features of the flow structures are in good agreement. The velocity profile at the exit of the packing near the walls as compared to centre of the packed bed shows a very good agreement between the simulations and the experiments. For the BCC packing, the axial velocities can become higher at the peripheral pores than at the central pores of the packed bed. The discrepancies between simulation and experiments may be attributed to the surface roughness of the 3D-printed spheres used in the experiments, as the behaviour of the boundary layers near the spheres is very sensitive to the surface roughness. This is corroborated  by the fact that fluctuations in the axial velocity are higher in the experiments compared to the velocities predicted by the simulations. Hence, the velocity field at the exit can be affected, especially in the case of the BCC packing, where spheres are closely packed, and therefore the boundary layers are more pronounced, than in the SUC packing. Another probable reason for the discrepancy is the relatively low spatial resolution achieved in the SPIV measurements and the subsequent underestimation of the volumetric flow rate by the SPIV measurements, compared to the volumetric flow rate as determined by the mass flow controllers. However, in general, both the experiments and the simulations show a good agreement, and the complex flow structures in the packings can be well predicted.

\section{Acknowledgements}
\noindent 
This work was funded by the Deutsche Forschungsgemeinschaft (DFG, German Research Foundation) - Project-ID 422037413 - TRR 287. Gef\"ordert durch die Deutsche Forschungsgemeinschaft (DFG) - Projektnummer 422037413 - TRR 287.\\
We would like to thank Dr. Gunar Boye for helping with the preparation of the experimental setup. We would like to thank M.Sc. Afrin Merchant for carrying out the 3D-printing of the packed bed units. 




\bibliographystyle{cas-model2-names}


\clearpage
\section*{Nomenclature}
\noindent
$A$ Area, m$^2$\\
$d$ Diameter, m\\
$F$ Force, N\\
$f$ Frequency, Hz\\
$g$ Gravitational acceleration, m/s$^2$\\
$m$ Mass, kg\\ 
$p$ Pressure, Pa\\
$Q$ Volumetric flow rate, m$^3$/s\\
$Re$ Reynolds number, -\\
$s$ Source term, - \\
$t$ Time, s\\
$u$ or $U$ Velocity, m/s\\
$V$ Volume, m$^3$\\
$W$ Spreading weight, -\\
$\mathbf{X}$ Lagrangian marker position, m\\
$\mathbf{x}$ Eulerian cell center position, m\\
$y$ Wall distance, m\\
$Z$ Axial direction\\
$X$ and $Y$ Lateral directions\\

\subsection*{Greek letters}
\noindent
$\gamma$ Fluid variable, -\\
$\Gamma$ Lagrangian variable, -\\
$\mu$ Dynamic viscosity, Pa*s\\
$\rho$ Density, kg/m$^3$\\
$\tau$ Stress tensor, N/m$^2$\\
$\phi$ Interpolation weight, -\\

\subsection*{Super- and subscripts}
\noindent
n - time level\\
j - j-th Lagrangian marker\\
IB - Immersed boundary\\
int - Interstitial\\
p - Particle\\
spf - Superficial\\
J - Central Jet\\
C - Co-flow\\

\subsection*{Abbrevations}
\noindent
BCC - Body-Centered Cubic packing\\
SUC - Simple unit cell packing\\
CFD - Computational fluid dynamics\\
CFL - Courant-Friedrichs-Lewy number\\
IBM - Immersed Boundary Method\\
LBM - Lattice-Boltzmann Method\\
MRI - Magnetic Resonance Imaging\\
PIV - Particle Image Velocimetry\\
SPIV - Stereo Particle Image Velocimetry\\
RT-PIV - Ray Tracing Particle Image Velocimetry\\
RIM - Refractive Index Matching\\
CCD - Charged Coupled Device\\
PTU - Programmable Time Unit\\
RMS - Root Mean Square\\
SLS - Selective Laser Sintering\\
DEHS - Di Ethyl Hexyl Sebacat\\

\end{document}